# Chloride ions as integral parts of hydrogen bonded networks in aqueous salt solutions: the appearance of solvent separated anion pairs†

Ildikó Pethes[*,a], Imre Bakó[b], László Pusztai[a,c]

[a]Wigner Research Centre for Physics, Konkoly Thege út 29-33., H-1121 Budapest, Hungary

[b]Research Centre for Natural Sciences, Magyar tudósok körútja 2, H-1117 Budapest, Hungary

[c]International Research Organisation for Advanced Science and Technology (IROAST), Kumamoto University, 2-39-1 Kurokami, Chuo-ku, Kumamoto 860-8555, Japan

**Abstract**

Hydrogen bonding to chloride ions has been frequently discussed over the past 5 decades. Still, the possible role of such secondary intermolecular bonding interactions in hydrogen bonded networks has not been investigated in any detail. Here we consider computer models of concentrated aqueous LiCl solutions and compute usual hydrogen bond network characteristics, such as distributions of cluster sizes and of cyclic entities, both for models that take and do not take chloride ions into account. During the analysis of hydrogen bonded rings, a significant amount of 'solvent separated anion pairs' have been detected at high LiCl concentrations. It is demonstrated that taking halide anions into account as organic constituents of the hydrogen bonded network does make the interpretation of structural details significantly more meaningful than when considering water molecules only. Finally, we compare simulated structures generated by 'good' and 'bad' potential sets on the basis of the tools developed here, and show that this novel concept is, indeed, also helpful for distinguishing between reasonable and meaningless structural models.

---

[*] Corresponding author: e-mail: pethes.ildiko@wigner.hu

†Electronic supplementary information (ESI) available. See DOI: …

**Introduction**

The term 'hydrogen bonding' (H-bonding) is traditionally connected to water and ice, where oxygen atoms form bonds with hydrogen atoms of neighboring water molecules by enhancing the electronic density between O and (non-bonded) H by the lone electron pairs of the O atoms[1-3]. The same mechanism works for many other compounds with hydroxyl (-OH) groups: examples are alcohols, organic acids, sugars, proteins, DNA, etc… However, hydrogen bonding is responsible for the very strong links between molecules of hydrogen fluoride, $HF$[4,5], too: that is, hydrogen bonds (HB-s) may be formed not only by oxygen-, but also, by halogen atoms/halide ions (and even by nitrogen, c.f. liquid ammonia, $NH_3$)[6].

The phenomenon of hydrogen bonding with halogen atoms/halide ions is, indeed, well documented in the literature: for instance, the crystal structure of hydrogen chloride hydrates has been investigated by diffraction methods already half a century ago[7,8]. A fairly large number of hydrogen bond distances to halide ions in crystals have been reported a couple of decades ago[9]. More recent, related investigations can be found in Refs. 10,11. Hydrogen bonding in chloride-water clusters has been considered by Xantheas[12], and the dynamics of HB-s have been studied, via molecular dynamics (MD) simulations, in aqueous solutions of halide ions[13].

However, quite surprisingly, according to the best of our knowledge, halide ions have never been considered as integral parts of the network of hydrogen bonds in aqueous solutions of, e.g., alkali halide salts. Indeed, most of the (faintly) related discussions in earlier papers have been about how ions actually break/disrupt the hydrogen bonding network of water molecules in these solutions (see, e.g., Refs. 14, 15, 16). The traditional 'structure making/structure breaking' roles of ions[17] also concern the issue whether the presence of ions enhances or deteriorates of the HB network of water molecules. What happens when hydrogen bonds between halide anions and water molecules are both treated as network formers has not yet been investigated.

There is a notable, qualitative difference between the ways chloride ions and O atoms of water molecules act as (electron) donors while forming hydrogen bonds. Water molecules provide extra electron density by the lone electron pairs of their oxygen atoms, i.e., via *localised* electrons. The extra electron (providing the 1- negative charge) of the chloride ions, on the other hand, is distributed evenly over the 'surface' of the ion. This difference is the main reason why water oxygens can donate electrons for two (at most three) neighbouring H-atoms, whereas chloride ions can form 6(-7) hydrogen bonds[18,19] (without specific orientations). In other words, the number of H-bonds formed by chloride ions is limited by steric effects only. A brief demonstration, employing quantum mechanical calculations, of the notion that the Cl…H-O connection has the attributes of a standard H-bond can be found in the Supplementary Information.

Here, we make use of some of the standard hydrogen bonding related analysis tools[20,21] applied recently to water-methanol[22], water-ethanol[23] and water-isopropanol[24] liquid mixtures. The main goal was to learn whether the 'water only' (WO) ('pure') or the 'chloride-included' (CI) ('mixed') approach provides the more sensible description of the hydrogen bonded network in concentrated aqueous LiCl solutions. For this reason, each descriptor of the H-bond network was determined both for the 'WO' and 'CI' situations. In what follows, a systematic comparison between WO and CI characteristics is provided.

**Computational section**

As all details of the computer simulations are identical to those that have already been published in details[18], only a short summary is provided here (a more complete account can be found in the Supplementary Information).

Classical molecular dynamics (MD) simulations, at 300 K in the NVT ensemble, were performed by the GROMACS software package (version 5.1.1)[25]. The calculations were performed at constant volume and temperature (NVT ensemble), at $T$ = 300 K. Cubic simulation boxes were used with periodic boundary conditions. Four different concentrations of aqueous LiCl solutions (from 3.74 mol/kg to 19.55 mol/kg) and pure water were investigated. Simulation boxes contained about 10000 atoms. Box lengths were calculated according to the experimental densities. The numbers of ions and water molecules, densities, and box sizes are collected in Table 1.

**Table 1.** Aqueous LiCl solutions investigated: number of ions and water molecules, densities and simulation box sizes. Experimental densities are taken from Ref. 26.

| Molality [mol/kg] | 0 | 3.74 | 8.30 | 11.37 | 19.55 |
|---|---|---|---|---|---|
| $N_{LiCl}$ | 0 | 200 | 500 | 700 | 1000 |
| $N_{water}$ | 3333 | 2968 | 3345 | 3416 | 2840 |
| Density [g/cm$^3$] | 0.9965 | 1.076 | 1.1510 | 1.1950 | 1.2862 |
| Number density [Å$^{-3}$] | 0.0999 | 0.09735 | 0.0939 | 0.0919 | 0.0871 |
| Box length [nm] | 4.6425 | 4.5721 | 4.8982 | 5.0232 | 4.94102 |

Potential parameters applied in this study were chosen from the collection of Ref. 19, where 29 force field models were compared according to their appropriateness for describing the structure of highly concentrated aqueous LiCl solutions. In the rest of this work, results obtained by using (one of) the best model(s), JC-S, a model of Joung and Cheatham, III [27], are presented. For comparison, data from a 'bad' model, RM, a force field set of Reif and Hünenberger[28] are also shown. Several other models have also been tested: their potential parameters and results obtained are presented in the Supplementary Information, along with a demonstration of the 'goodness-of-fit'-s with respect to the measured structure factors[26].

During the simulations water molecules were kept rigid by the SETTLE algorithm[29]. Coulomb interactions were treated by the smoothed particle-mesh Ewald (SPME) method[30,31], using a 10 Å cutoff in direct space. The van der Waals interactions were also truncated at 10 Å, with added long-range corrections to energy and pressure[32].

Initial particle configurations were obtained by placing ions and water molecules randomly into the simulation boxes. Energy minimization was carried out using the steepest descent method. After that the leap-frog algorithm was applied for integrating Newton's equations of motion, using a time step of 2 fs. The temperature was kept constant by the Berendsen thermostat[33] with $\tau_T$=0.1 coupling. After a 4 ns equilibration period, particle configurations were collected at every 80 ps between 4 and 12 ns. The 101 configurations thus obtained were used for hydrogen bond analyses.

Hydrogen bonds (HB) can be identified by several methods. Results presented here have been obtained by applying geometric considerations[34]. All calculations were repeated using the energetic definition of HB-s[34]: findings of that are shown in the Supplementary Information. According to the geometric definition, two water molecules are identified as H-bonded if the intermolecular distance between an oxygen and a hydrogen atom is less than 2.5 Å, and the O...O-H angle is smaller than 30 degrees. A chloride ion is considered H-bonded to a water molecule if the H...Cl$^-$ distance is less than 2.8 Å and the Cl$^-$...O-H angle is smaller than 30 degrees. According to the energetic definition, in addition to the criteria on the above O...H (Cl$^-$...H) distance, the interaction energy between H-bonded molecules (molecule and ion) should be less than -3.0 kcal/mol.

Determination of H-bonded molecules and calculations concerning the H-bonded network were performed by an in-house programme, based on the HBTOPOLOGY code.[20]

Before moving on to displaying and interpreting our present results, a short note is perhaps appropriate here concerning the possibility of applying not classical, but quantum computer simulations ('*ab initio* molecular dynamics, AIMD'). Unfortunately, to the best of our knowledge, no relevant AIMD calculation has been performed for highly concentrated aqueous LiCl solutions. What is available for this electrolyte is a simulation with 64 water molecules and two ion pairs,[35] which means a concentration of about 3 molar % -- whereas the lowest concentration in the present study is about twice that. Also, no direct comparison with measured structural data in the reciprocal space can be found in the corresponding AIMD papers and therefore, it is not possible to establish the relevance of the structural information derived from these high level simulation studies. Although this statement is valid for the thorough AIMD study of Gaiduk et al.[16] on a (rather dilute) aqueous solution of NaCl, there is an issue in this AIMD work that is relevant from the point of view of the present investigation. Gaiduk et al. discuss the effect of ions on the hydrogen bonded network of water molecules, and they seem to have had a difficulty with interpreting the role of chloride ions in this respect. What we show in the following is that such a difficulty transforms into a sensible explanation when chloride ions are considered not as modifiers, but *organic constituents* of the network of H-bonds in such solutions.

**Results and discussion**

We wished to facilitate comparability with literature by taking exactly the same salt concentrations as in previous, related works from our laboratory[15,18,19,26], namely 3.74 m (mol/kg; corresponding to 6.3 molar %), 8.3 m (13 molar %), 11.37 m (17 molar %) and 19.55 m (26 molar %). In fact, the calculations providing the atomic assemblies ('particle configurations') were identical to those reported recently by one of us[18]. Note that these concentration values are rather high, the highest one representing an ion/water ratio of about 50:75. The reason why such systems have been selected was the expectation that the role of ions in enhancing/disrupting the hydrogen bonded network would be most apparent under such circumstances.

Simulation details, as well as a brief description the geometric definition of hydrogen bonds used here throughout are provided in the 'Computational section' (see above, and in the Supplementary Information). Most importantly, H-bonded $Cl^-…H$ maximum distances are somewhat longer than $O…H$ ones, as it is defined by the first minimum of the $Cl^-$-H and O-H partial radial distribution functions (see, e.g., Refs.[15,18,19]). The potential model called 'JC-S' from Ref. 19 is used just below for demonstrating features of the 'mixed' H-bond network concept.

**Cluster size distributions**

In order to characterize the extent of the H-bonded network, cluster size distributions (as defined in, e.g. Refs. 20,21) have first been determined; these are shown in Figure 1. Robust hydrogen bonded networks, like those in most alcohol-water mixtures, percolate[21,23,24], i.e. the largest H-bonded cluster is comparable in its size with the system size. From Figure 1 it is obvious that when chloride ions are not considered as 'network formers' then this criterion is fulfilled only for pure water and, to some extent, for the least concentrated LiCl solution. In the more concentrated solutions one can only find isolated water clusters, up to sizes of about 180 (11.37 m) or even only about 20 (19.55 m) molecules. That is, the H-bonded networks of water molecules are really small in comparison with the system size and also, such a picture would suggest a kind of 'microphase separation', debated quite hotly in the cases of alcohol-water mixtures[36,37]. On the other hand, when chloride ions are taken into account as parts of the network then cluster sizes are equal (within a few percent) with the cumulative number of water molecules and chloride ions.

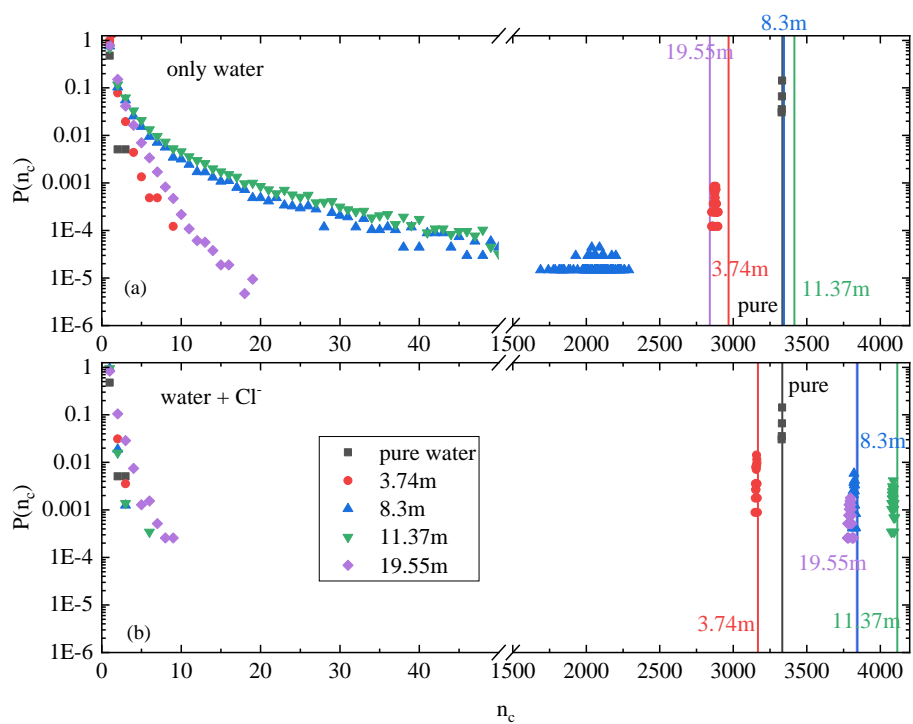

**Figure 1.** Cluster size distributions calculated for the same atomic configurations. (a) 'WO', water molecules only (no Cl⁻ ions in the H-bonded network). (b) 'CI', water molecules AND chloride ions. (The vertical lines show the number of (a) water molecules (b) the Cl⁻ ions plus water molecules in the system.) Note that when chloride ions are included, even the most concentrated systems percolate – which makes sense in a homogeneous solution.

**Hydrogen bonded rings**

Next, the occurrence of H-bonded cyclic entities is scrutinized (Figure 2). The number or purely water rings decreases dramatically with increasing ion concentration, with hardly any cycles present above 8.3 m. On the other hand, there is a fair amount of 'mixed' cycles even at the highest ion concentration when the anions are also counted (even though the number of H-bonded cycles decreases also here when salt concentration is growing). Interestingly, the size of the rings also decreases with increasing concentration, so much, that the most frequent ring size in the 19.55 m LiCl solution contain only 4 members.

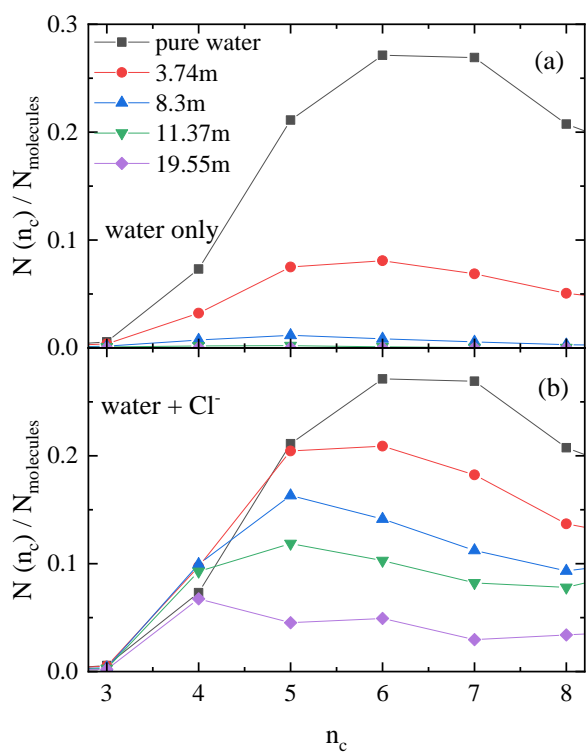

**Figure 2.** Size distributions of cyclic entities, as calculated for the same particle configurations, but (a) 'WO', without chloride ions, and (b) 'Cl', with chloride ions in the H-bonded network. Note the trend: the number of rings decreases, and rings become smaller as salt concentration increases.

It is also instructive to investigate the ratio of water molecules and chloride ions in the 'mixed' cycles (Figure 3). Contrary to what was observed in methanol-water liquid mixtures[22], the participation of ions in the H-bonded rings follows roughly the overall concentration of (an)ions in the solutions. Again, it is the most concentrated solution that exhibits the most spectacular feature: the most frequent cycle is the one that consists of two water molecules and two chloride ions (see Figure 4 for representative parts of the particle configurations).

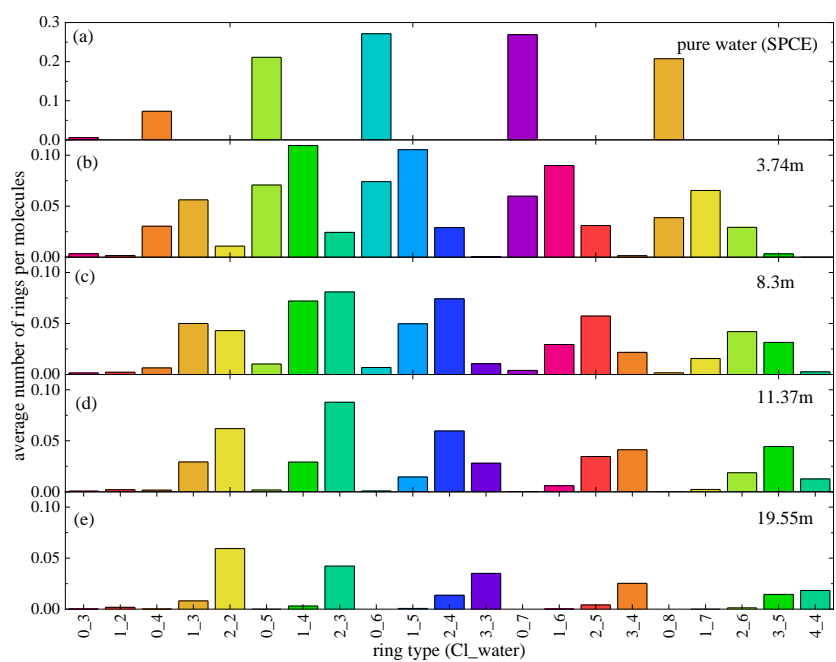

**Figure 3.** Distribution of different types of rings (rings contain Cl$^-$ ions and water molecules), normalized by the number of molecules (water + Cl$^-$ ions), at different concentrations obtained from the JC-S model. Note that the scale of part (a) is different from the others.

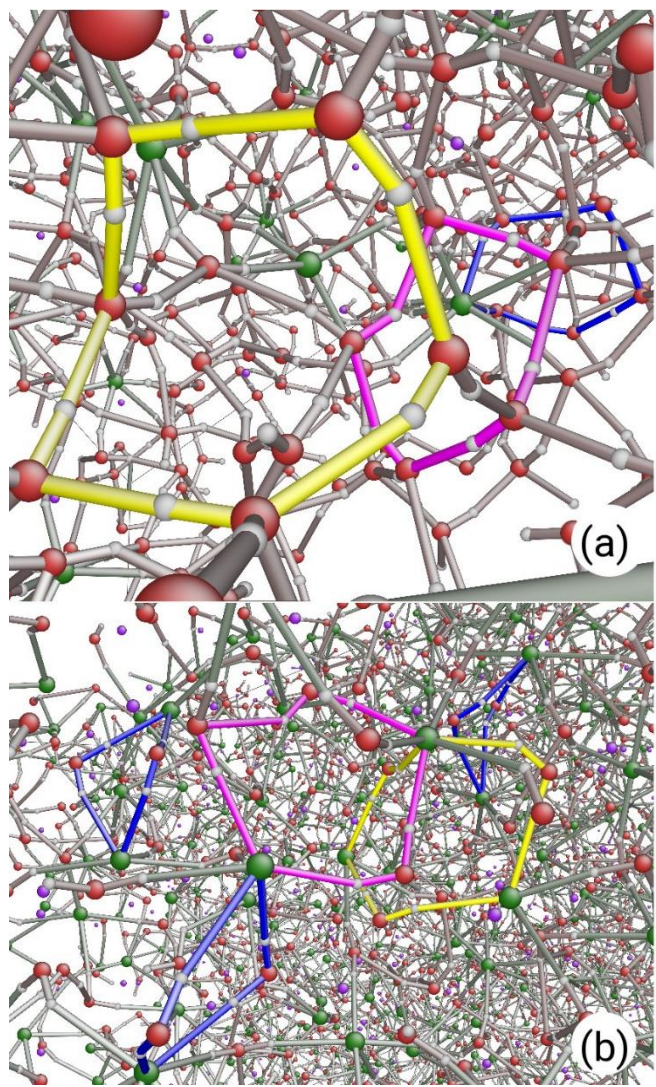

**Figure 4.** Snapshots of configurations obtained with JC-S forcefield at the concentrations (a) $m$=3.74 and (b) $m$=19.55 mol/kg. Red, gray, purple and green balls represent oxygen, hydrogen, lithium and chlorine atoms, respectively. Some frequent ring types are highlighted, on part (a): (blue) 6-membered ring with 1 $Cl^-$ ion and 5 water molecules, (magenta) 5-membered ring with 5 water molecules and (yellow) 6-membered ring contains 6 water molecules; on part (b) (blue) 4-membered rings with 2 $Cl^-$ ion and 2 water molecules, (magenta) 5-membered ring with 2 $Cl^-$ ions and 3 water molecules and (yellow) 6-membered ring contains 3 $Cl^-$ ions and 3 water molecules.

The commonsense expectation is that no two chloride ions would be connected directly (it would not even be any kind of a 'hydrogen bond'). Indeed, closer inspection reveals (cf. Figure 4, part (b)) that a frequently occurring constellation is where there is one water molecule between two chloride ions: these motifs can be considered as 'solvent separated anion pairs'. Even though the presence of such particle arrangements is not entirely unexpected (at least once the 'mixed' water/anion H-bonded network concept has been introduced), to our best of knowledge, this is the first occasion when this phenomenon is detected in simulation models, and pictured in a very straightforward manner.

Two quick tests have been performed concerning these solvent separated anion pairs: (1) it has been verified that in a given solution, the H-bonding energy (calculated according to Refs. [21, 23, 24]) of a $Cl^-…H-O$ hydrogen bond does not depend on whether this H-bond is a single one, or part of a solvent separated anion pair; (2) the lifetime (calculated according to Ref. [24]) of solvent separated anion pairs in a given solution is actually about two times that of a single (and also, solitary) $Cl^-...H-O$ bond. More details can be found in the Supplementary Information.

So far, it has been shown that there are marked differences between the concepts of considering the 'pure' hydrogen bonded network of water molecules only, and that of a 'mixed' network that includes chloride ions, too. We believe that the latter provides a more sensible characterization of a homogeneous liquid – and since no sign of any small angle scattering could be spotted on either the neutron or the X-ray data[26] we argue that however concentrated the solutions in question are, even the 19.55 m LiCl solution is homogeneous.

**Utilization of the concept**

Next, we further demonstrate the usefulness of the 'mixed' concept by comparing 'good' and 'bad' potential models for aqueous LiCl solutions (cf. Ref. 19). As 'good' force field, the JC-S combination from Ref. 19 was taken, that consists of the SPC/E water model[38] and the ionic parameters from Ref. 27. The 'bad' example, called RM in Ref. 19, contains SPC/E water molecules combined with ions as represented in Ref. 28. The difference between the two force field combinations is that while the JC-S one reproduces experimental neutron and X-ray diffraction data very well, the RM combination fails to do so. Further details concerning potential parameters are provided in the 'Computational section'.

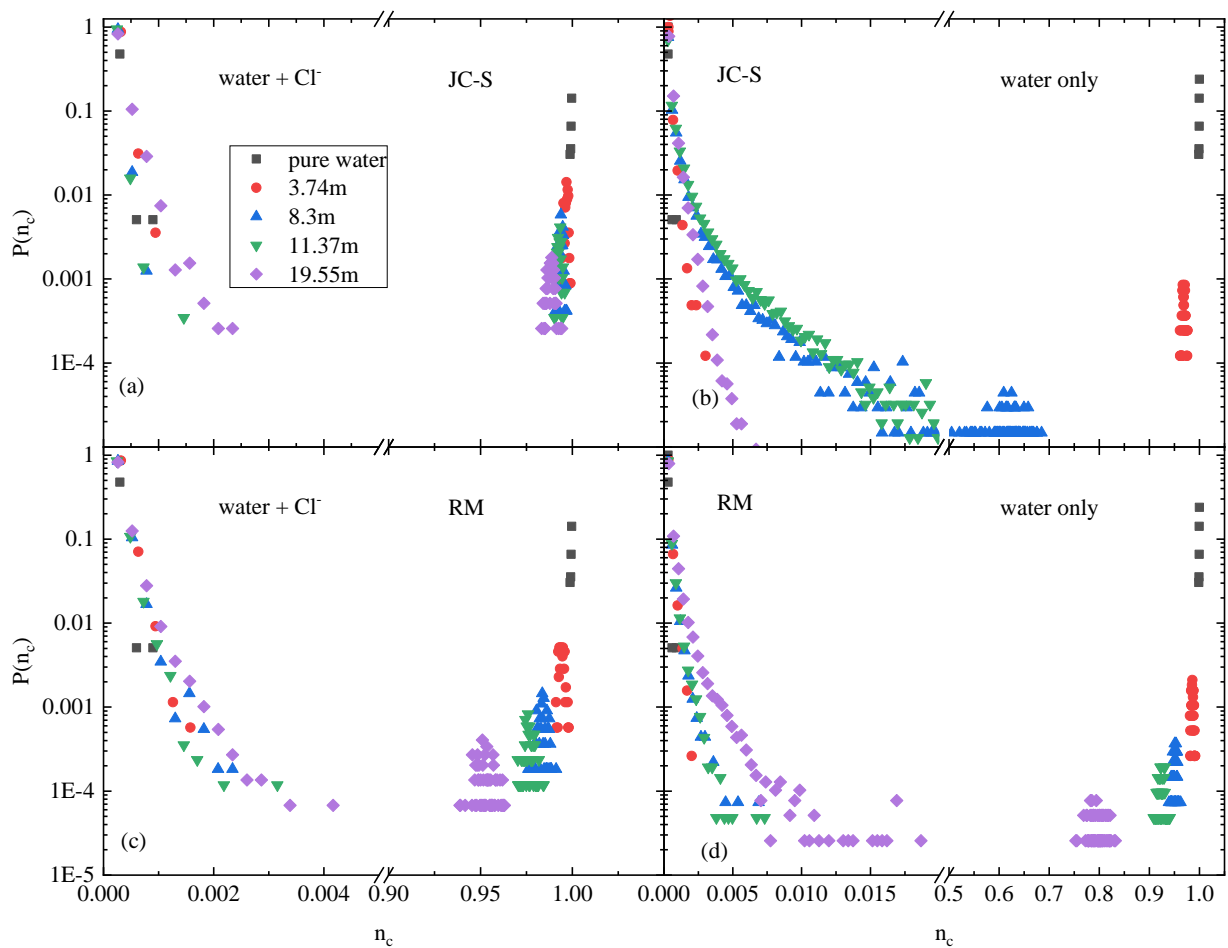

**Figure 5.** Cluster size distributions calculated for the atomic configurations obtained from (a, b) JC-S and (c, d) RM models. (a, c) Water molecules AND chloride ions. (b, d) Water molecules only (no Cl$^-$ ions in the H-bonded network). (The x-axes are normalized with the number of (a, c) Cl$^-$ ions plus water molecules, (b, d) water molecules in the configurations.) Note that for the RM model, which was found to be one of the worst when comparing simulated and experimental total structure factors, cluster sizes with Cl$^-$ ions (part (c)) appear to be smaller than the percolation limit, whereas pure water clusters (part (d)) are large even at high salt concentrations. This indicates non-perfect mixing (i.e., 'microphase-segregation' between water and salt), which is against the observation that all solution considered are homogeneous.

Figure 5 shows cluster size distributions for the JC-S and RM models, using both the 'pure water' and 'mixed' definitions of the H-bonded network. (Note that the presentation is more condensed here than it was in Figure 1: on the '*x*' axis, cluster size values are shown as normalized to the system size.) The difference between the two potential combinations is striking: while pure water H-bonded clusters are larger, mixed water-anion ones are smaller in the case of the 'bad' RM model. These observations are consistent with the notion that mixing of ions and water is far from perfect when the RM combination of force fields is applied, leading to a kind of (micro-)phase segregation, between water-rich and ion-rich regions. Please remember: the RM model cannot reproduce diffraction data appropriately, i.e. the behavior detected in Figure 5 for this model cannot be related to characteristics of the real system.

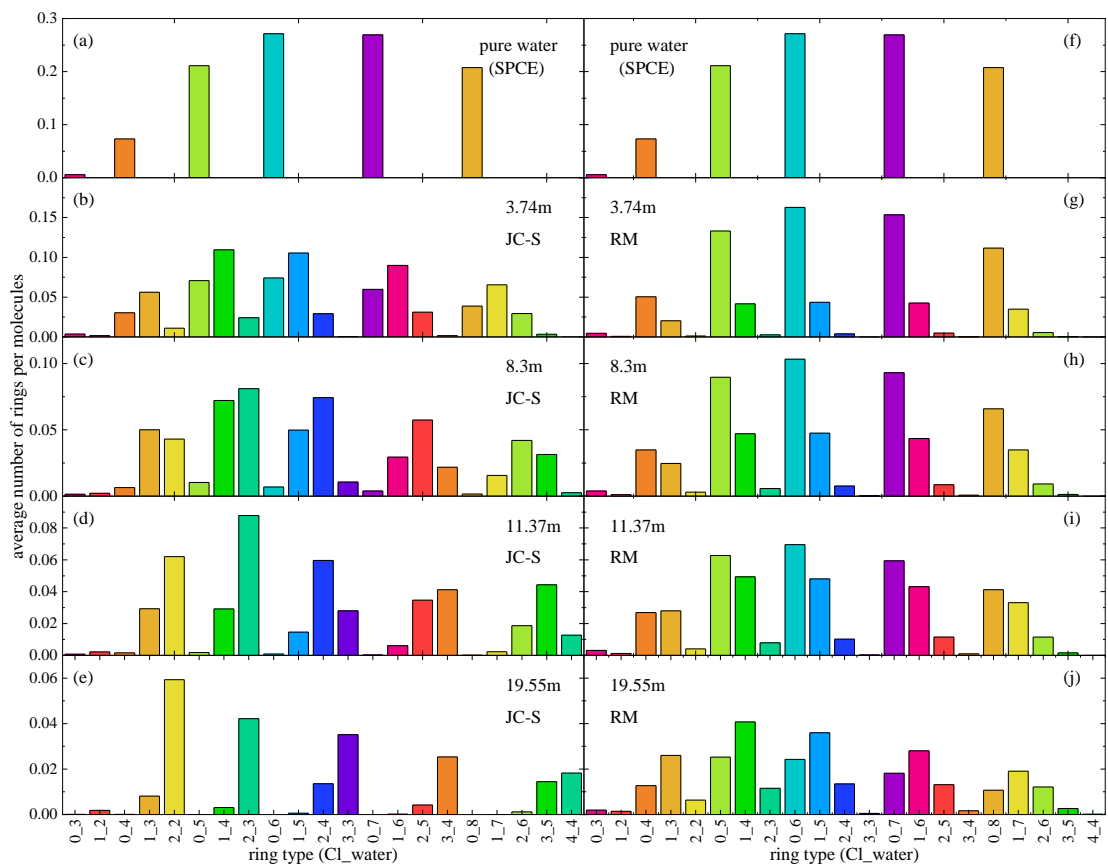

**Figure 6.** Distribution of different types of rings (rings contain Cl$^-$ ions and water molecules), normalized by the number of molecules (water + Cl$^-$ ions), at different concentrations, obtained from (b-e) JC-S and (g-j) RM models. The ring size distribution in pure water (a,f) is also shown for reference. Note the difference between the number of pure water rings (dominating in the 'bad' RM model, part (g-j)). Again, this shows the (inappropriate, if good solutions are present) tendency against mixing Cl$^-$ ions and water molecules in the 'bad' model.

When the composition of cyclic entities formed in the JC-S and RM structural models is compared (see Figure 6), an analogous conclusion can be drawn: while in the 'good' JC-S structure H-bonded rings contain a number Cl$^-$ ions that is in accord with the overall concentration of ions, cycles in the 'bad' RM structure tend to contain far less anions than it could be expected from the ion concentration in the solutions. In the RM model, the overwhelming majority of rings is water-dominated even at the highest ionic concentration. This, again, is an indication that the RM combination of water and ion force fields does not lead to homogeneous structures.

## Conclusions

In summary, it has been demonstrated that the concept of 'mixed' water-anion hydrogen bonded network provides a sensible characterization of highly concentrated chloride salt solutions. Percolating HB networks, with the participation of chloride ions, could be identified even in the most highly concentrated solution. The 'mixed' anion-water network can account for the homogeneity of such systems, contrary to what the 'pure' water network suggests. The concept may well appear to be helpful for interpreting earlier findings concerning hydrogen bonding networks of aqueous halide salt solutions (c.f., e.g., Refs. 14, 15, 16).

Concerning hydrogen bonded cyclic entities, the novel concept reveals that the participation of chloride anions in rings is proportional to the ionic concentration. Cycles containing 2 water molecules and 2 chloride ions have been found to be the dominant motifs at the highest salt concentration.

The approach has brought about the observation of 'solvent separated anion pairs' that are the dominant motifs in cyclic hydrogen bonded entities at high LiCl concentration (see Figures 3 and 4).

The characterization of, as well as the distinction between, 'good' and 'bad' potential models of aqueous LiCl solutions becomes very natural via the 'mixed' network concept: good models facilitate mixing of ions and water molecules at the atomic scale, whereas inappropriate force fields tend to result in separation of solvent and solute (micro-)phases (cf. Figures 5 and 6).

Further explorations are needed (and underway) for establishing whether the 'mixed' water+halide ion hydrogen bonded network is a useful concept in general for discussing properties of highly concentrated aqueous solutions of (at least) alkali-halides. As a possible next step, we will first look at a situation where the counter-ion is the largest of alkali cations, namely the case of CsCl solutions (for which experimental data is available from our group[39]).

## Conflicts of interest

There are no conflicts to declare.

## Acknowledgments

The authors are grateful to the National Research, Development and Innovation Office (NKFIH) of Hungary for financial support through Grants Nos. 124885 and KH130425.

# Supplementary Information

# Chloride ions as integral parts of hydrogen bonded networks in aqueous salt solutions: the appearance of solvent separated anion pairs

Ildikó Pethes[a], Imre Bakó[b], László Pusztai[a,c]

[a]Wigner Research Centre for Physics, Konkoly Thege út 29-33., H-1121 Budapest, Hungary

[b]Research Centre for Natural Sciences, Magyar tudósok körútja 2, H-1117 Budapest, Hungary

[c]International Research Organisation for Advanced Science and Technology (IROAST), Kumamoto University, 2-39-1 Kurokami, Chuo-ku, Kumamoto 860-8555, Japan

Corresponding author: e-mail: pethes.ildiko@wigner.hu

# I. On the ionic/covalent character of Cl⁻…H-O hydrogen bonds: quantum chemical considerations

Hydrogen bonding (H-bonding) is an important interaction that plays a key role in chemical, physical, and biochemical processes. H-bonding is significantly weaker than a typical chemical bond, but stronger than van der Waals interactions.

From the point of view of the present work, it would be essential to know (and demonstrate) that the interaction between chloride anions and water molecules, indeed, show characteristics of hydrogen bonds. Although quite some heuristic arguments are mentioned in the main text, we wished to provide further evidence in support of the claim. In this small demonstration, we use the 'Atom in Molecule' (AIM) [1 - 4] and 'Natural Bond Orbitals' (NBO) [5 - 8] approaches for characterizing the chloride ion - water interaction.

Bader's theory [1] of atoms-in-molecules (AIM) is an elegant theoretical tool for understanding both covalent and non-covalent molecular interactions. Within the framework of this theory, one can investigate topological properties of the electron density in the molecule. One of the most important feature in this theory the existence of the bond critical point along a bond. It has already been shown that there is a correlation between the strength of the hydrogen bonding interaction and the properties at this point (charge density, $\rho$; ellipticity, ...). In a weak/medium hydrogen bond the charge density is in the range of 0.002..0.04 and the Laplacian of the electron density, $\Delta\rho$, is positive (0.02..0.13), when using the definition of Popelier et al. [3].

The Natural Bond Orbitals method provides us with a deeper insight of the electron transfer process from the lone pair of a Lewis base to an unfilled OH* antibonding orbital of a Lewis acid. This type of interaction can stabilize 'complexes', like the hydration shells of chloride ions. Both the AIM and NBO based calculations of the real charge on the chloride anion may quantitatively yield a hint on the strength/covalent character of an H-bond.

The geometries of Cl⁻...(H$_2$O)$_6$ surface clusters were optimized at the M052x/cc-pvtz level of theory, which method is suitable for such calculations [9]. The initial cluster geometries were taken from the literature [10 - 12]. Properties of three different inner shell clusters were also investigated. The clusters considered here are shown in Figure S1. The DFT geometry optimization and single point energy calculations were performed by utilizing the Gaussian09E.01 [13] program package. For reference, some properties of water dimers have also been computed by the same method.

**Table S1**. AIM and NBO properties of the clusters (see Figure S1) considered.

|  | $\rho$ (AIM) | $\Delta\rho$ (AIM) | ellipticity (AIM) | Charge of the Cl$^-$ ion AIM/NBO |
|---|---|---|---|---|
| water dimer | 0.026 | 0.082 | 0.028 | |
| Cl$^-$...6 water surface | 0.023 (±0.006) | 0.052 (±0.006) | 0.021 (±0.006) | -0.82/-0.85 |
| Cl$^-$...6 water inner shell | 0.016 (±0.002) | 0.045 (±0.003) | 0.047 (±0.02) | -0.80/-0.87 |

Table S1 contains the aforementioned AIM and NBO properties of the investigated clusters. It is clear from the both the AIM and NBO charges of Cl$^-$ that there is a significant charge transfer in these complexes shown in Figure S1. Also note that values for the other properties shown are in the range of the corresponding values calculated for water dimers.

That is, it may be concluded that the Cl$^-$…H-O interaction is certainly not purely electrostatic but shows all characteristics of a 'standard' hydrogen bond.

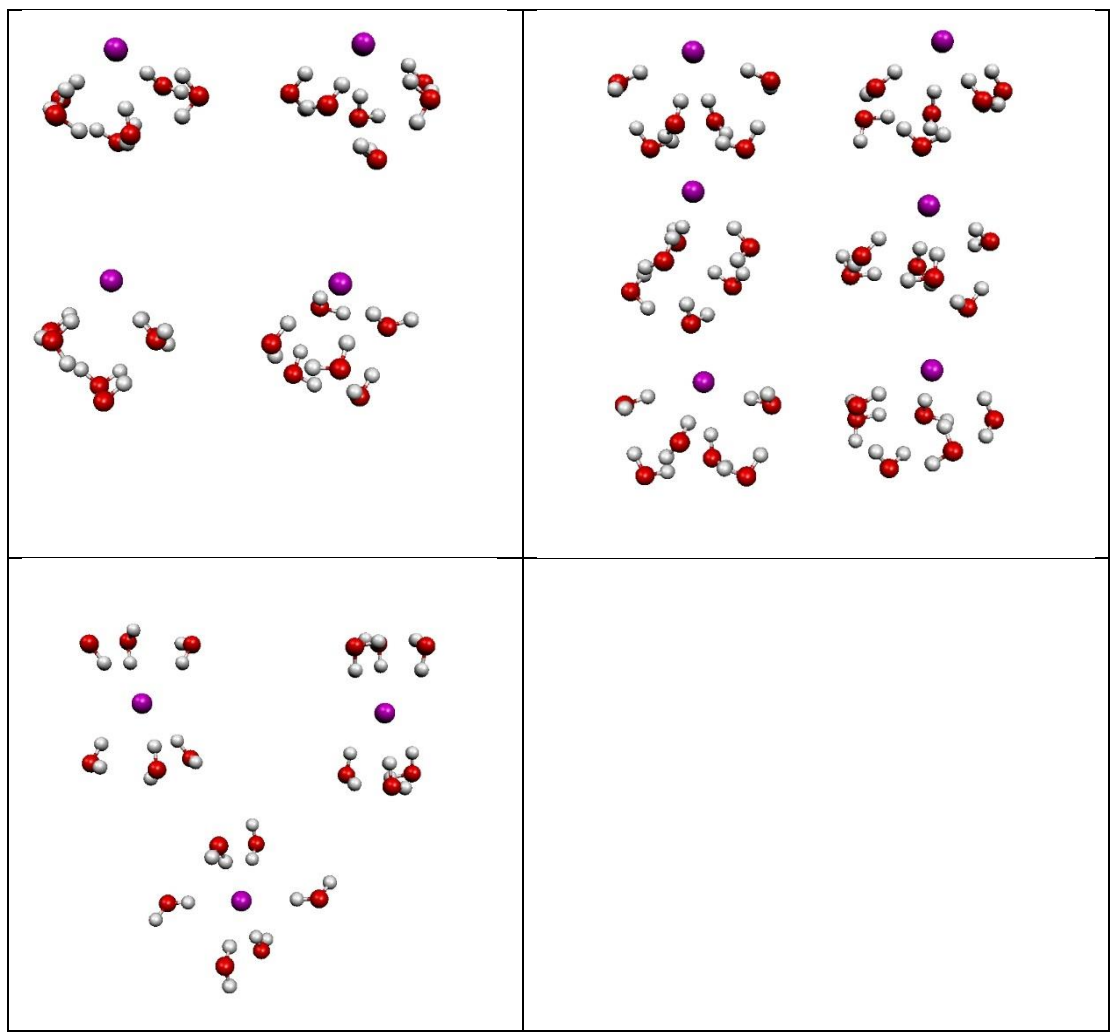

**Figure S1.** Clusters of Cl⁻ ions with 6 water molecules considered in this demonstration. Upper panels: anion on the surface; lower panel: anion in the inner shell.

## II. Molecular dynamics simulations

Classical molecular dynamics (MD) simulations were performed by the GROMACS software package (version 5.1.1) [14]. The calculations were performed at constant volume and temperature (NVT ensemble), at $T = 300$ K. Cubic simulation boxes were used with periodic boundary conditions. Four different concentrations of aqueous LiCl solutions (from 3.74 mol/kg to 19.55 mol/kg) and pure water were investigated. The simulation boxes contained about 10000 atoms, the box lengths were calculated according to the experimental densities. The number of ions, water molecules, densities and box sizes are collected in Table 1 of the main text.

Pairwise additive non-polarizable intermolecular potentials was applied for the description of interatomic interactions. The non-bonded interactions are described by the Coulomb potential accounting for electrostatics and the 12-6 Lennard-Jones (LJ) potential for the van der Waals interactions:

$$V_{ij}(r_{ij}) = \frac{1}{4\pi\varepsilon_0}\frac{q_i q_j}{r_{ij}} + 4\varepsilon_{ij}\left[\left(\frac{\sigma_{ij}}{r_{ij}}\right)^{12} - \left(\frac{\sigma_{ij}}{r_{ij}}\right)^{6}\right]. \tag{1}$$

Here $r_{ij}$ is the distance between particles $i$ and $j$, $q_i$ and $q_j$ are the point charges of the particles, $\varepsilon_0$ is the vacuum permittivity, $\varepsilon_{ij}$ and $\sigma_{ij}$ are the 12-6 LJ potential parameters. Potential parameters applied in this study were chosen from the collection of Ref. [15], in which paper 29 force field models were compared according to their appropriateness to describe the structure of highly concentrated aqueous LiCl solutions.

The $q_i$, $\varepsilon_{ii}$ and $\sigma_{jj}$ parameters of six tested models are collected in Table S2, the parameters of the corresponding water models are shown Table S3. The $\varepsilon_{ij}$ and $\sigma_{ij}$ values (parameters between unlike atoms) are calculated according to the Lorentz-Berthelot (LB) or the geometric (geom) combination rule, also shown in Table S2. In the geometric combination rule, both the $\varepsilon_{ij}$ and $\sigma_{ij}$ are calculated as the geometric average of the homoatomic parameters, whilst in the Lorentz-Berthelot type the $\varepsilon_{ij}$ is calculated as geometric, and $\sigma_{ij}$ as the arithmetic average of the relevant parameters.

**Table S2.** Force field parameters of the potential models investigated. The applied water models and combination rules are also shown. For the definitions of the combination rules, see the corresponding text.

| Model | $q_{Li}/q_{Cl}[e]$ | $\sigma_{LiLi}$ [nm] | $\varepsilon_{LiLi}$ [kJ/mol] | $\sigma_{ClCl}$ [nm] | $\varepsilon_{ClCl}$ [kJ/mol] | comb. rule | water model | References |
|---|---|---|---|---|---|---|---|---|
| JC-S | +1/-1 | 0.1409 | 1.4089 | 0.4830 | 0.0535 | LB | SPC/E | [16] |
| JC-T | +1/-1 | 0.1440 | 0.4351 | 0.4918 | 0.0488 | LB | TIP4PEw | [16] |
| MP-T | +1/-1 | 0.1715 | 0.2412 | 0.4612 | 0.1047 | LB | TIP4PEw | [17] |
| AqCh | +1/-1 | 0.2126 | 0.0765 | 0.4417 | 0.4928 | geom | SPC/E | [18, 19] |
| RM | +1/-1 | 0.3078 | 0.0015 | 0.3771 | 1.1137 | geom | SPC/E | [20] |
| JJ | +1/-1 | 0.2870 | 0.0021 | 0.4020 | 2.9706 | geom | TIP4P | [21] |

**Table S3.** Parameters of the water-models. In the TIP4P and TIP4PEw models there is a fourth (virtual) site (M). It is situated along the bisector of the H-O-H angle and coplanar with the oxygen and the hydrogen atoms. The negative charge is allocated to M.

| | $\sigma_{OO}$ [nm] | $\varepsilon_{OO}$ [kJ/mol] | $q_H$ [e] | $d_{O-H}$ [nm] | $\theta_{H-O-H}$ [deg] | $d_{O-M}$ [nm] | Ref. |
|---|---|---|---|---|---|---|---|
| SPC/E | 0.3166 | 0.6502 | +0.4238 | 0.1 | 109.47 | - | [22] |
| TIP4P | 0.3154 | 0.6485 | +0.52 | 0.09572 | 104.52 | 0.015 | [23] |
| TIP4PEw | 0.3164 | 0.6809 | +0.52422 | 0.09572 | 104.52 | 0.0125 | [24] |

According to Ref. [15] the models' appropriateness to describe the structure of the highly concentrated aqueous LiCl solutions is more or less proportional to the number of the contact ion pairs predicted by the model. The tested six potential models were selected to cover the full range of the numbers of the contact ion pairs found in Ref. [15] (see Table S4).

**Table S4.** Average numbers of the contact ion pairs ($N_{LiCl}$ coordination numbers) predicted by the force field models investigated in Ref. [15].

| forcefield | 3.74m | 8.3m | 11.37m | 19.55m |
|---|---|---|---|---|
| JC-S | 0.01 | 0.08 | 0.24 | 1.28 |
| JC-T | 0.08 | 0.31 | 0.59 | 1.47 |
| MP-T | 0.46 | 0.93 | 1.27 | 1.86 |
| AqCh | 1.28 | 1.67 | 1.83 | 2.15 |
| RM | 1.95 | 2.26 | 2.32 | 2.51 |
| JJ | 2.05 | 2.26 | 2.35 | 2.49 |

In the main text results obtained by using (one of) the best model(s), JC-S, a model of Joung and Cheatham, III [16], are presented. For comparison, data from a 'bad' model, RM, a force field set of Reif and Hünenberger [20] is also discussed. Figure S2 demonstrates the difference between 'good' and 'bad' potential models. The 'JC-S' combination reproduces measured data, in the reciprocal space, at an at least semi-quantitative level: this combination is therefore used as 'good' combination in the main text.

During the simulations water molecules were kept together rigidly by the SETTLE algorithm [25]. Coulomb interactions were treated by the smoothed particle-mesh Ewald (SPME) method [26,27], using a 10 Å cutoff in direct space. The van der Waals interactions were also truncated at 10 Å, with added long-range corrections to energy and pressure [28].

Initial particle configurations were obtained by placing the ions and water molecules randomly into the simulation boxes. Energy minimization was carried out using the steepest descent method. After that the leap-frog algorithm was applied for integrating Newton's equations of motion, using a 2 fs time step. The temperature was kept constant by the Berendsen thermostat [29] with $\tau_T$=0.1 coupling. After a 4 ns equilibration period, particle configurations were collected in every 80 ps between 4 and 12 ns. The obtained 101 configurations were used for hydrogen bond analyses.

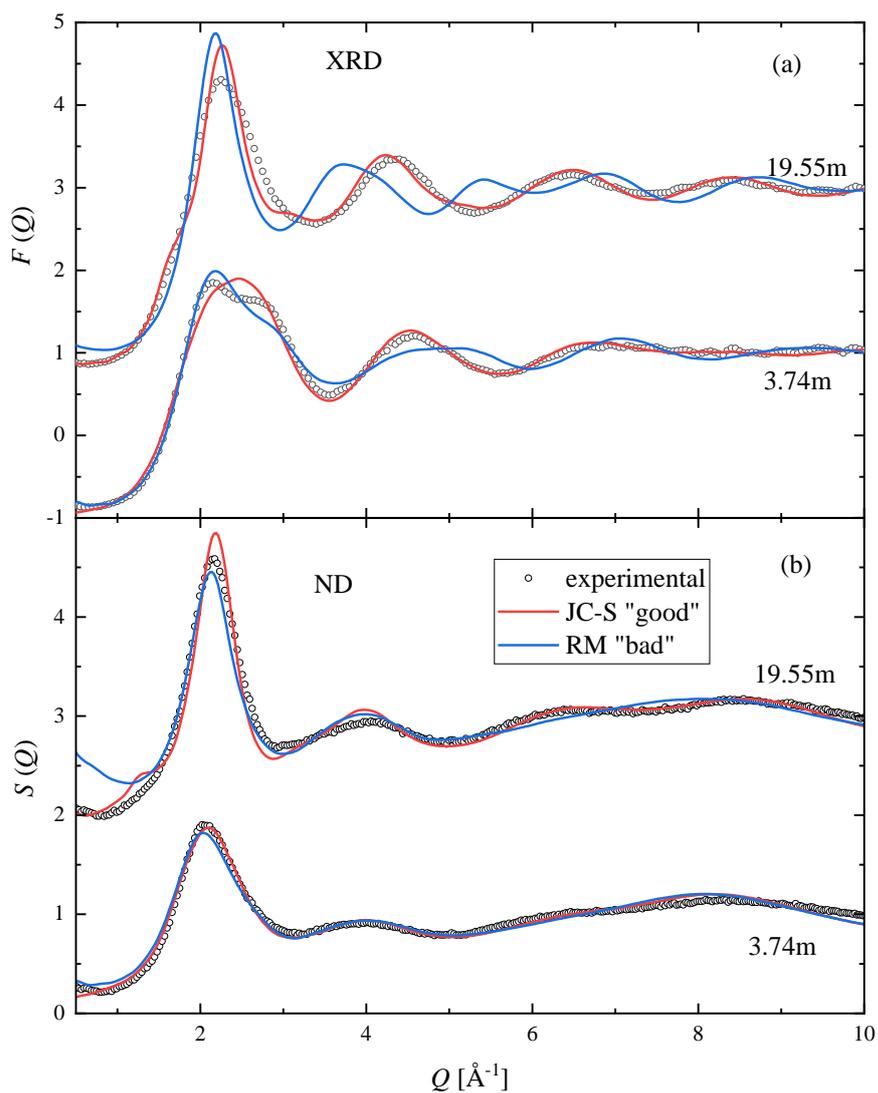

**Figure S2**. (a) X-ray and (b) neutron total structure factors from experiments (symbol, Ref. [30] ) and MD simulations using the JC-S (red) and the RM (blue) models for the 3.74m and 19.55m solutions.

# III. Cluster size distributions: 'proof of concept' – energetic definition of hydrogen bonds

In sections III., IV., VII., and VIII. we show the same kinds of graphs as Figures 1, 2, 3, 5 and 6 of the main text, but using the energetic definition of hydrogen bonds (see, e.g., Refs. [31 - 33]), instead of the purely geometric one that is utilized in the main text.

In short, the energetic definition of H-bonds works in conjunction with one simple geometric criterion as follows: two water molecules, as well as a chloride ion and a water molecule, are considered to be hydrogen-bonded to each other if they are found within a distance r(O…H / Cl$^-$…H) < 2.5 Å, and the interaction energy is smaller than -12 kJ/mol (ca. -3 kcal/mol).

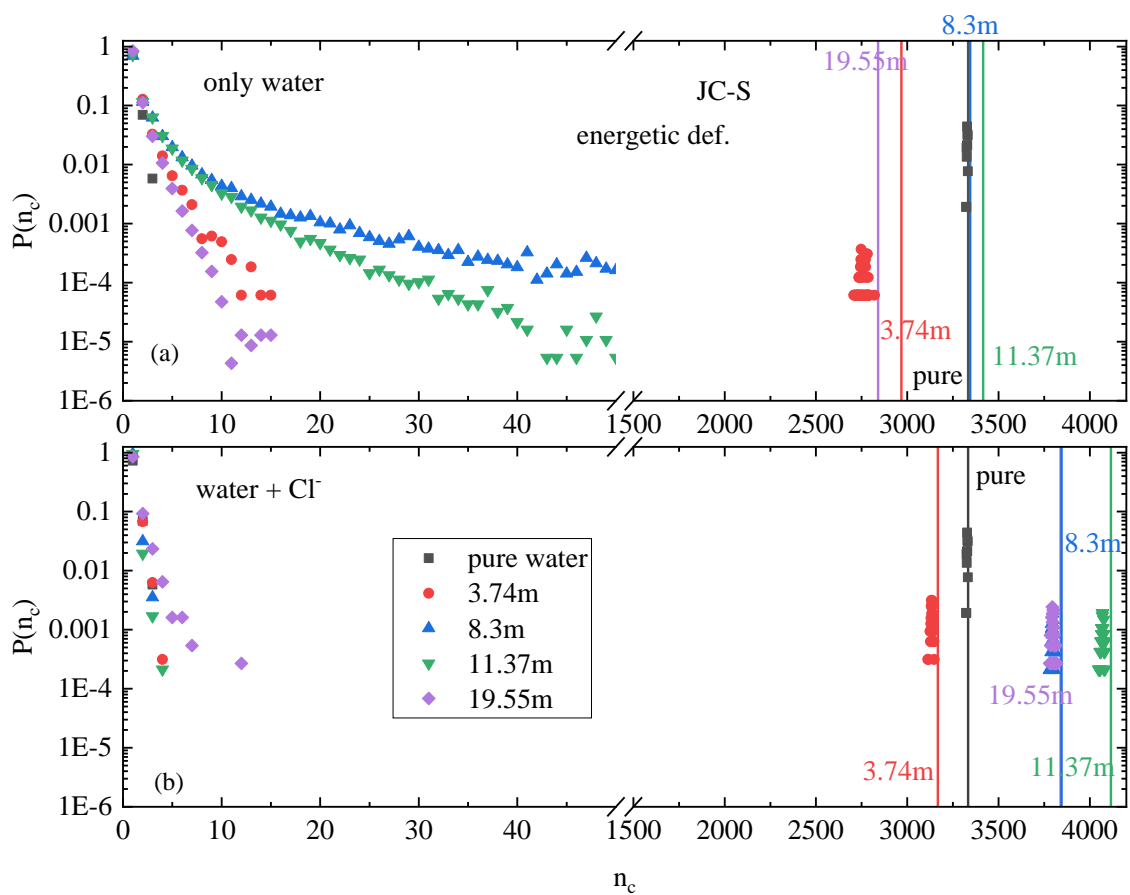

**Figure S3**. Cluster size distributions calculated for the same set of atomic configurations. (a) Water molecules only (no Cl$^-$ ions in the H-bonded network). (b) Water molecules AND chloride ions. (The vertical lines show the number of (a) water molecules (b) the Cl$^-$ ions plus water molecules in the system.) The JC-S potential model and the energetic definition of H-bond is used.

# IV. Cyclic entities: 'proof of concept' – energetic definition of hydrogen bonds

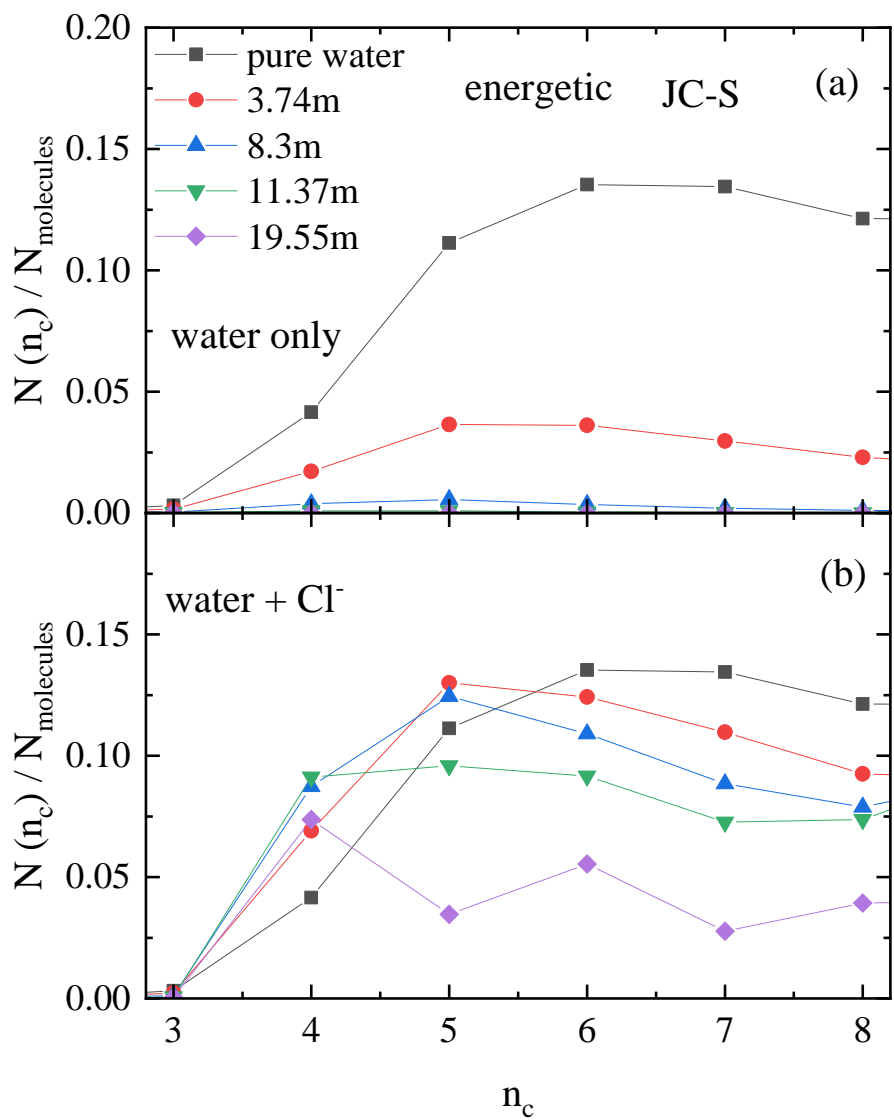

**Figure S4.** Size distribution of cyclic entities, as calculated for the same set particle configurations, but (a) without chloride ions, and (b) with chloride ions in the H-bonded network. The energetic definition of H-bond is used.

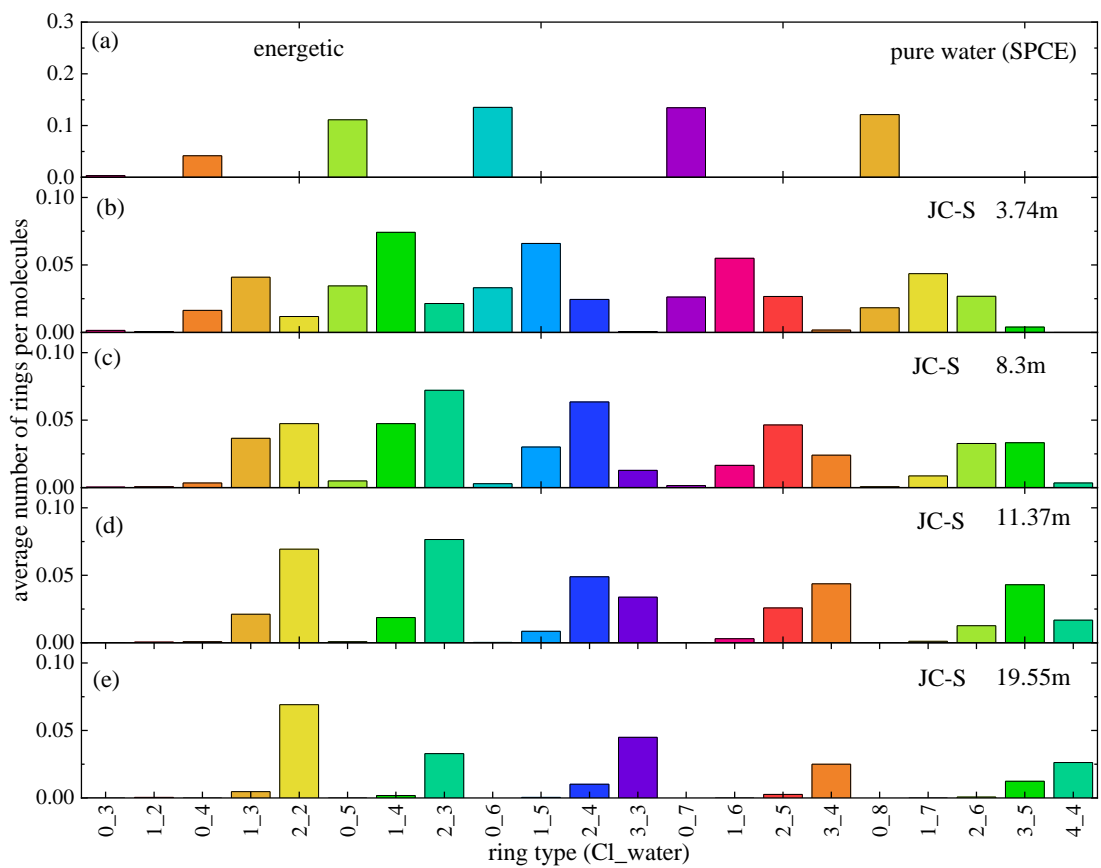

**Figure S5.** Distribution of different types of rings (rings contain Cl⁻ ions and water molecules), normalized by the number of molecules (water + Cl⁻ ions), at different concentrations obtained from JC-S model using the energetic definition of H-bond. Note: scaling in part (a) is different from that in the other parts.

# V. Hydrogen bond energetics

Bonding energies for single Cl$^-$…O-H bonds, as well as for those participating in solvent separated anion pairs have been calculated the same way as H-bond energies in, e.g. Refs. [32, 33].

Figure S6 contains the main findings of such calculations: at a given concentration, there is no difference between the energies of single Cl$^-$...O-H bonds and the ones that are parts of solvent separates anion pairs. On the other, energies of Cl$^-$-related H-bonds are getting somewhat deeper as LiCl concentration grows.

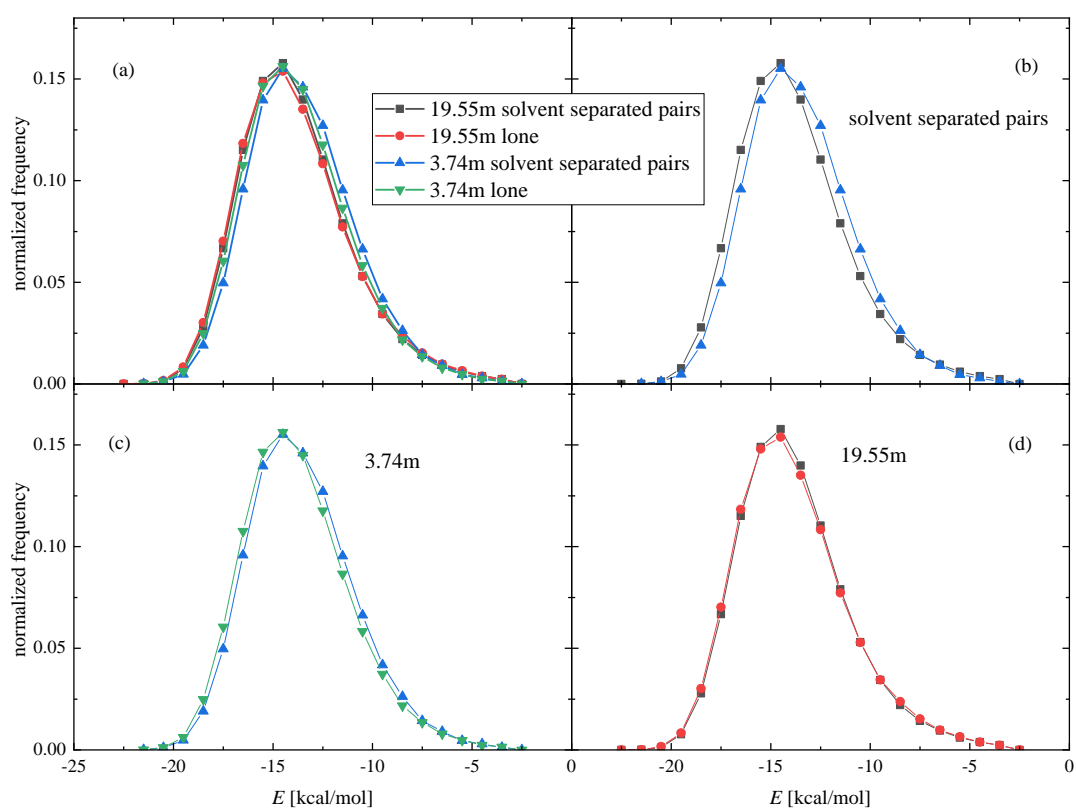

**Figure S6.** (a) Comparison of the hydrogen bond energies of Cl$^-$ - water pairs in solvent separated anion pairs (in structural motifs in which the water molecules have two H-bonded Cl$^-$ ion pairs) and in lonely Cl$^-$ - water pairs (in motifs in which the water molecules have only one H-bonded Cl$^-$ ion pair). (b) Comparison of the H-bond energies of Cl$^-$ - water pairs in solvent separated anion pairs at different concentrations. (c) and (d) Comparison of the hydrogen bond energies of Cl$^-$ - water pairs in solvent separated anion and in lonely Cl$^-$ - water pairs in the (c) 3.74m and (d) 19.55m solutions.

# VI. On the lifetimes of solvent separated anion pairs

We have studied the surviving probability (lifetime of H-bonds, as well as of the 'solvent-separated anion pairs') similarly as in Ref. [33], calculated according to the following function [9, 34, 35]:

$$c_n = \frac{\langle \delta h_n^I(t) \delta h_n^I(0) \rangle}{\langle \delta h_n^I(0) \delta h_n^I(0) \rangle} \quad (2)$$

where

$$\delta h_n^I(t) = h_n^I(t) - \langle h_n^I(t) \rangle \quad (3)$$

The function $h_n^I(t)$ has been defined in the following way:

$$h_n^I(t) = 1 \quad (4)$$

if a chloride ion or water molecule that was in the HB state *n* at time *t* = 0 is in the same HB state at time *t*, irrespective of whether or not its HB state has changed in the meantime, and 0 otherwise.

An estimate for the lifetime of a given arrangement (single HB, or solvent-separated anion pair) from this correlation function can be obtained by the following formula[34, 35]:

$$\tau_n^I = \int c_n^I \, dt \quad (5)$$

Table S5 shows that the solvent separated anion pairs actually live (a little more 2 times) longer than single Cl⁻...H-O hydrogen bonds. Another thing to note is that lifetimes grow drastically as the salt concentration approaches saturation (above 19.55 m).

**Table S5** Lifetimes (in picoseconds) of single Cl⁻…H-O hydrogen bonds and of Cl⁻…H-O-H…Cl⁻ solvent separated anion pairs in two of the aqueous LiCl solutions investigated here.

|  | lifetime (ps) | lifetime (ps) |
|---|---|---|
|  | 8.3 m | 19.55 m |
| Cl⁻...H-O | 7.60 | 433.52 |
| Cl⁻...H-O-H...Cl⁻ | 19.39 | 907.84 |

Results for pure liquid water at 298 K, using the TIP4P/2005 model, may be taken as reference, see Table 4 and Fig. 11 of Ref. [33]. As the chloride ion related H-bonds are much longer lived, it is evident also from this comparison that the Cl$^-$…H-O hydrogen bond is stronger than the O...H-O one present in pure water.

# VII. Cluster size distributions: comparison of 'good' and 'bad' models – energetic definition of hydrogen bonds

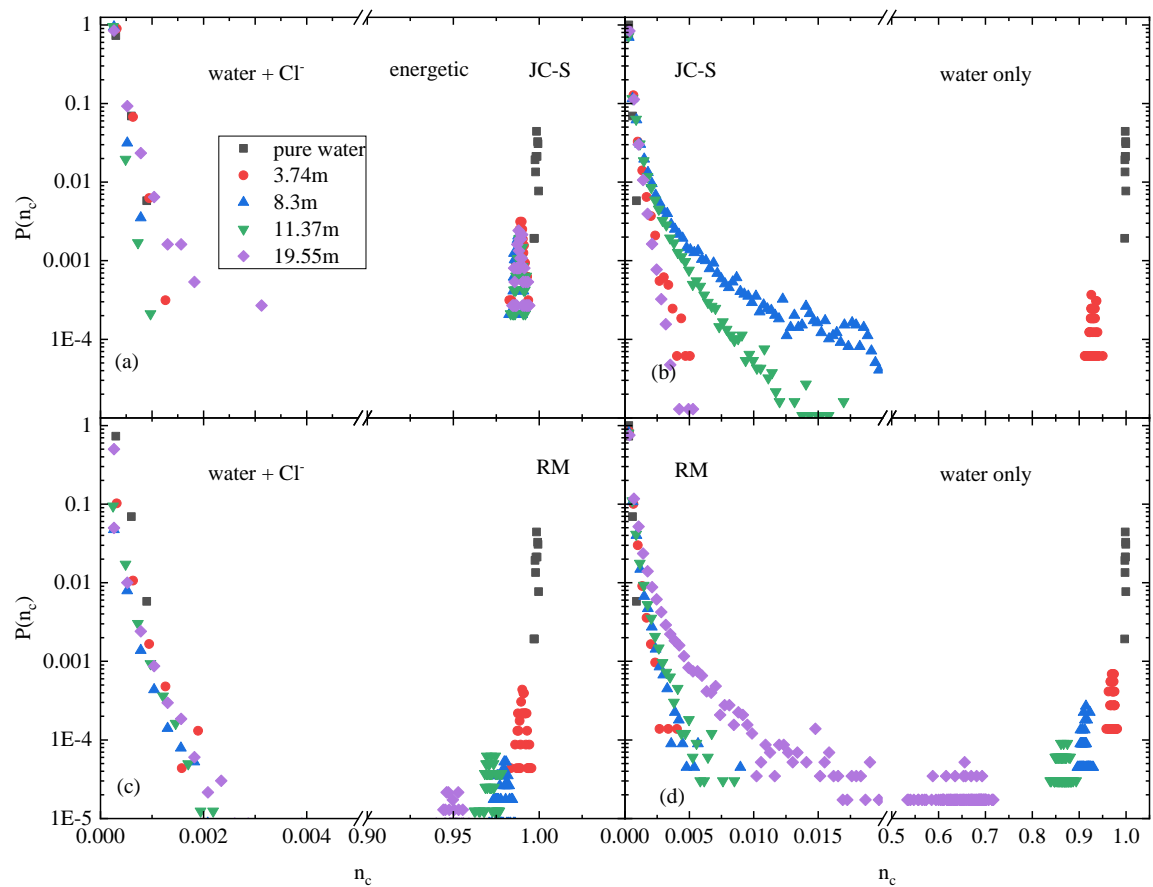

**Figure S7.** Cluster size distributions calculated for the atomic configurations obtained from (a, b) JC-S and (c, d) RM models. (a, c) Water molecules AND chloride ions. (b, d) Water molecules only (no Cl$^-$ ions in the H-bonded network). (The x-axes are normalized by the cumulative numbers of (a, c) Cl$^-$ ions plus water molecules, (b, d) water molecules in the configurations.) The energetic definition of H-bond is used.

# VIII. Cyclic entities: comparison of 'good' and 'bad' models – energetic definition of hydrogen bonds

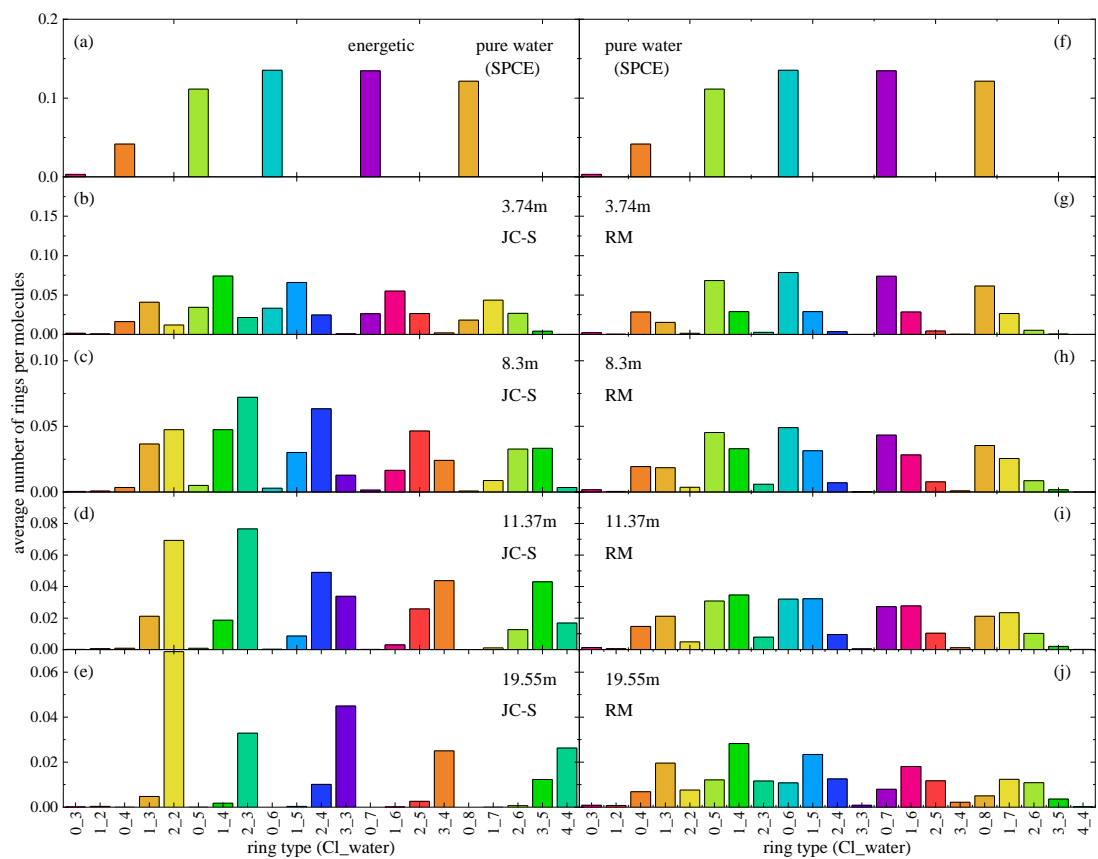

**Figure S8.** Distribution of different types of rings (rings contain Cl$^-$ ions and water molecules), normalized by the number of molecules (water + Cl$^-$ ions), at different concentrations, obtained from (b-e) JC-S and (g-j) RM models, using the energetic definition of H-bond. The ring size distribution in pure water (a,f) is also shown for reference.

# IX. Cluster size distributions: results for a range of potential models

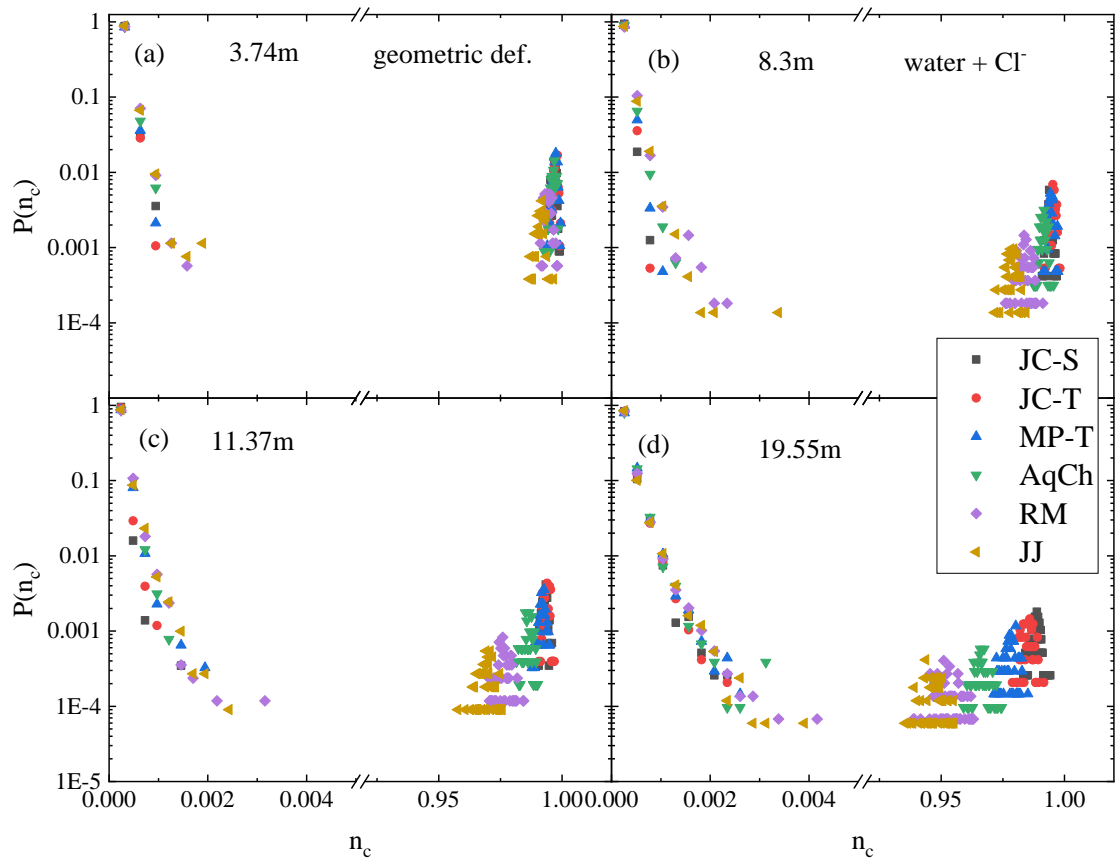

**Figure S9.** Cluster size distributions at various concentrations ( (a) 3.74 mol/kg, (b) 8.3 mol/kg, (c) 11.37 mol/kg, (d) 19.55 mol/kg) obtained from different MD models (using the geometric definition of H-bond). Water molecules AND chloride ions both are considered as parts of the network. (The x-axes are normalized by the cumulative numbers of Cl$^-$ ions plus water molecules in the configurations.)

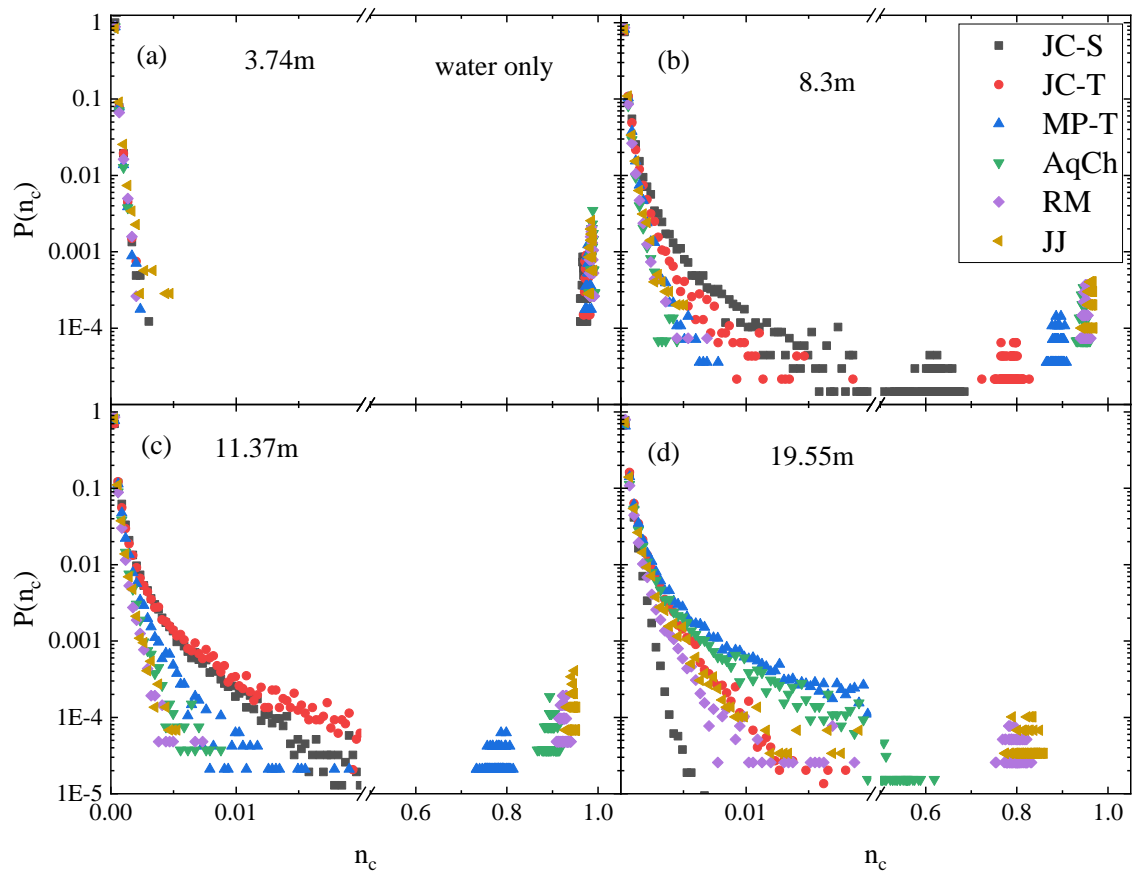

**Figure S10.** Cluster size distributions at various concentrations ( (a) 3.74 mol/kg, (b) 8.3 mol/kg, (c) 11.37 mol/kg, (d) 19.55 mol/kg) obtained from different MD models (using geometric definition of H-bonded molecules) Only water molecules are considered. (The x-axes are normalized by the numbers of water molecules in the configurations.)

# X. Cyclic entities: results for a range of potential models

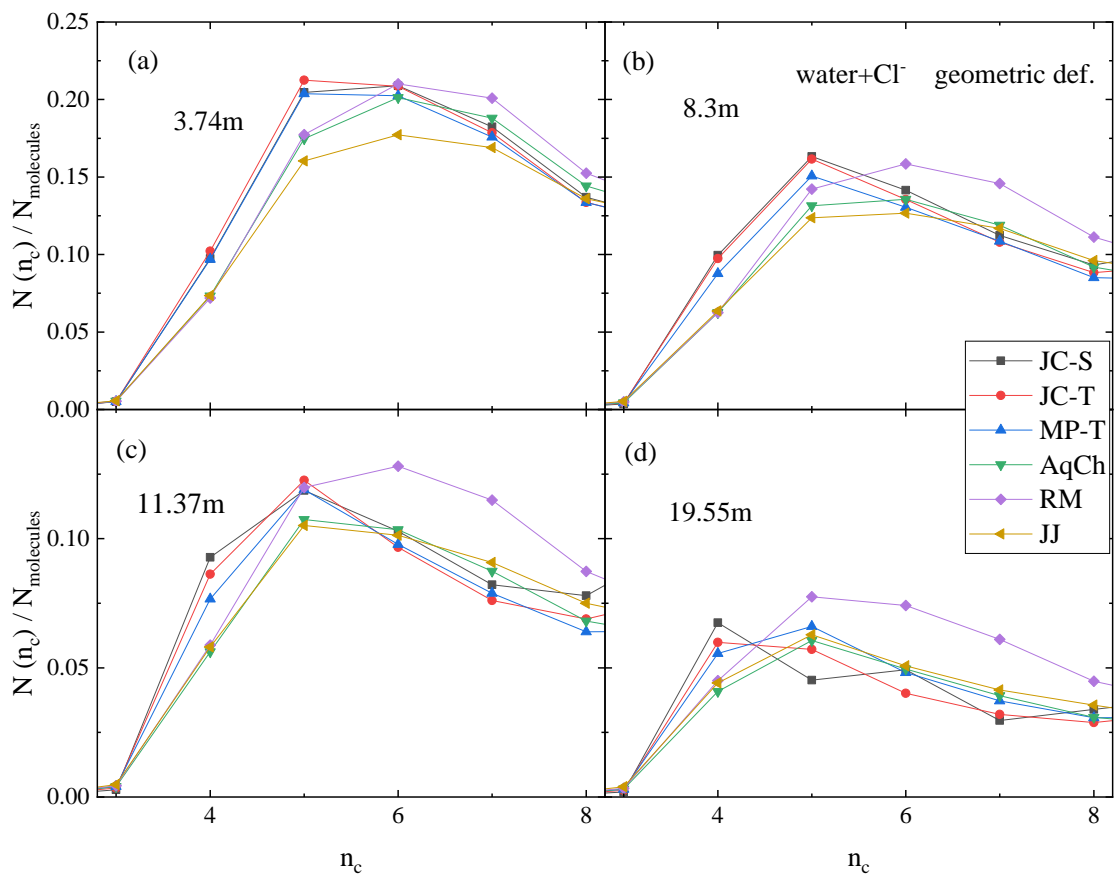

**Figure S11.** Size distributions of cyclic entities calculated with chloride ions in the H-bonded network at various concentrations ( (a) 3.74 mol/kg, (b) 8.3 mol/kg, (c) 11.37 mol/kg, (d) 19.55 mol/kg), obtained from different MD models (using the geometric definition of H-bond).

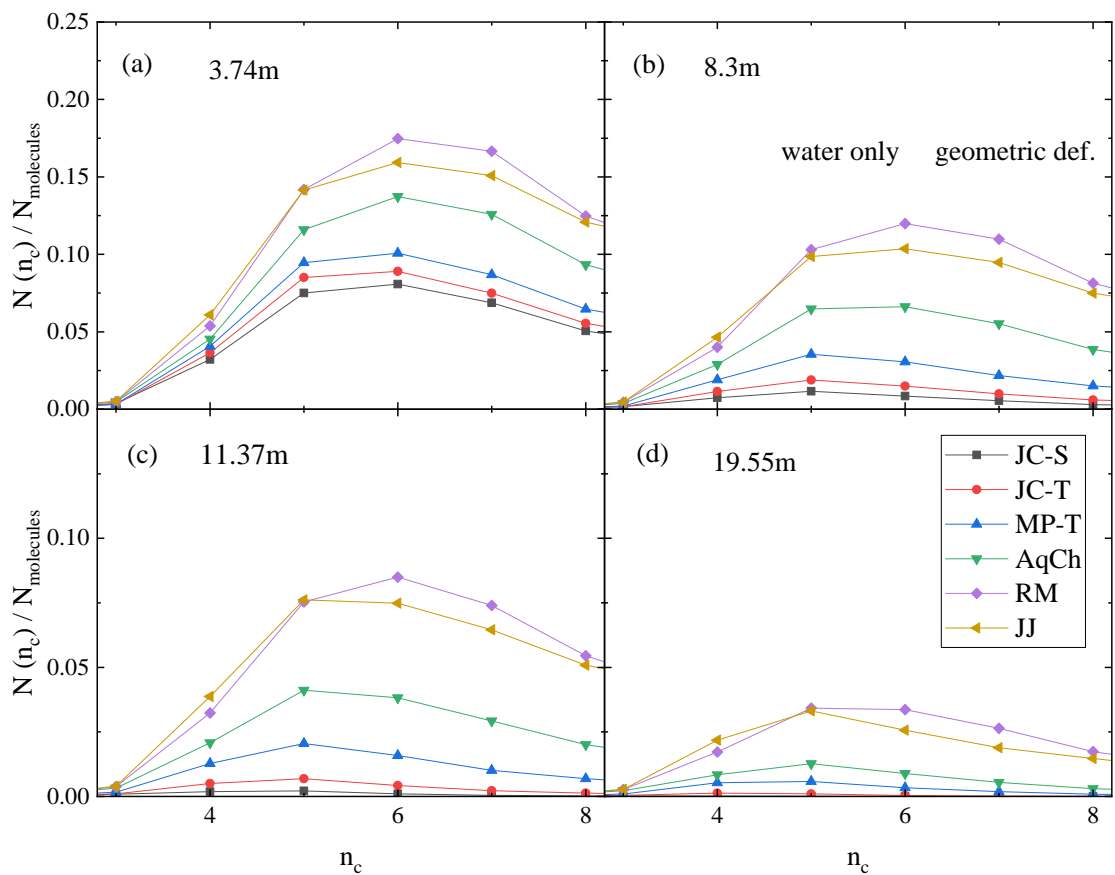

**Figure S12.** Size distribution of cyclic entities calculated without chloride ions in the H-bonded network, at various concentrations ( (a) 3.74 mol/kg, (b) 8.3 mol/kg, (c) 11.37 mol/kg, (d) 19.55 mol/kg), obtained from different MD models (using the geometric definition of H-bond).

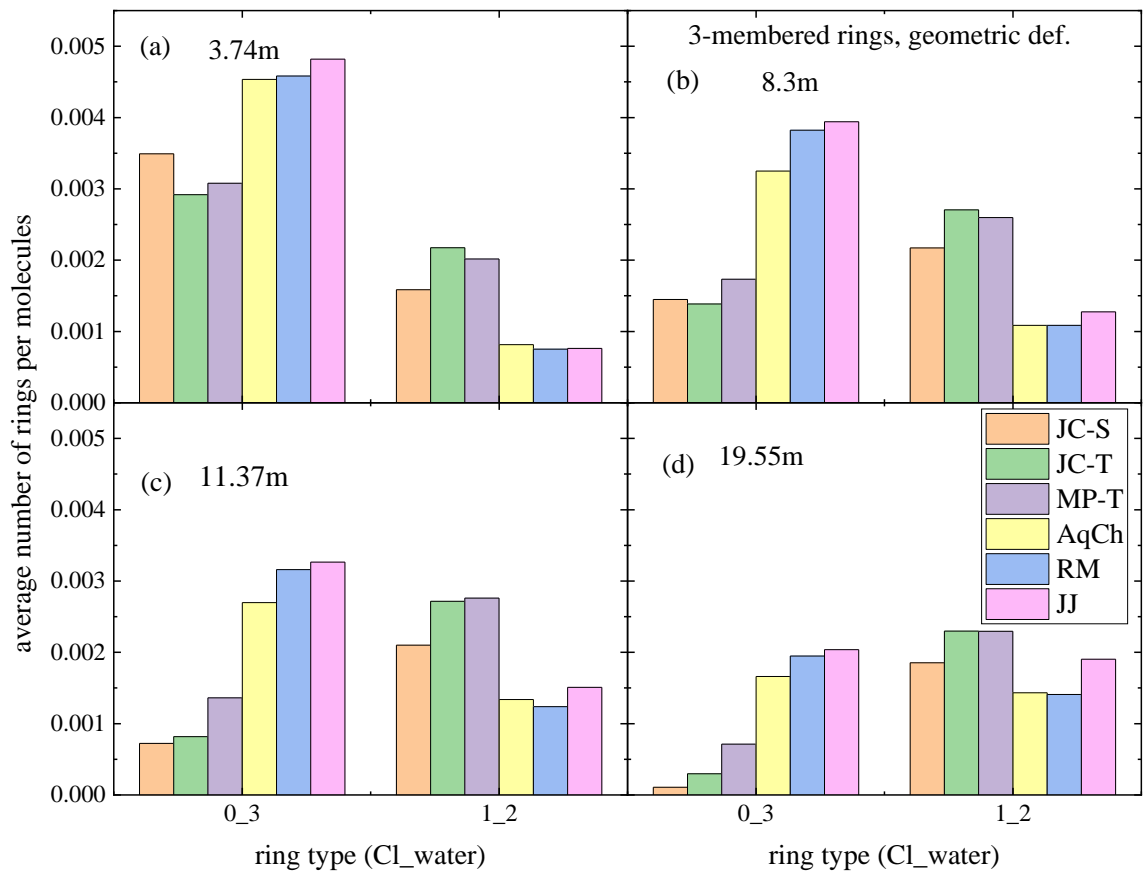

**Figure S13.** Distributions of different types of 3-membered rings (rings contain Cl⁻ ions and water molecules), normalized by the number of molecules (water + Cl⁻ ions), at different concentrations ( (a) 3.74 mol/kg, (b) 8.3 mol/kg, (c) 11.37 mol/kg, (d) 19.55 mol/kg), obtained from different MD models, using the geometric definition of H-bond.

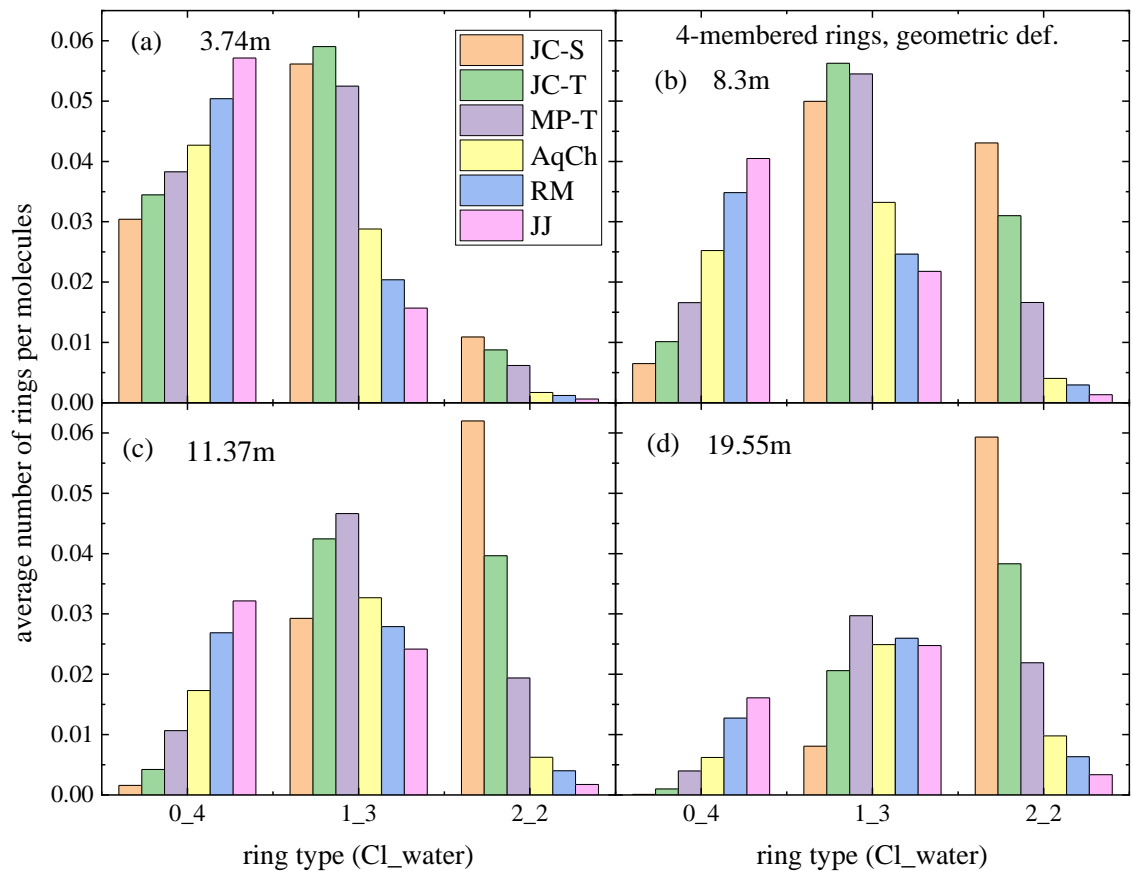

**Figure S14.** Distributions of different types of 4-membered rings (rings contain Cl$^-$ ions and water molecules), normalized by the number of molecules (water + Cl$^-$ ions), at different concentrations ( (a) 3.74 mol/kg, (b) 8.3 mol/kg, (c) 11.37 mol/kg, (d) 19.55 mol/kg), obtained from different MD models, using the geometric definition of H-bond.

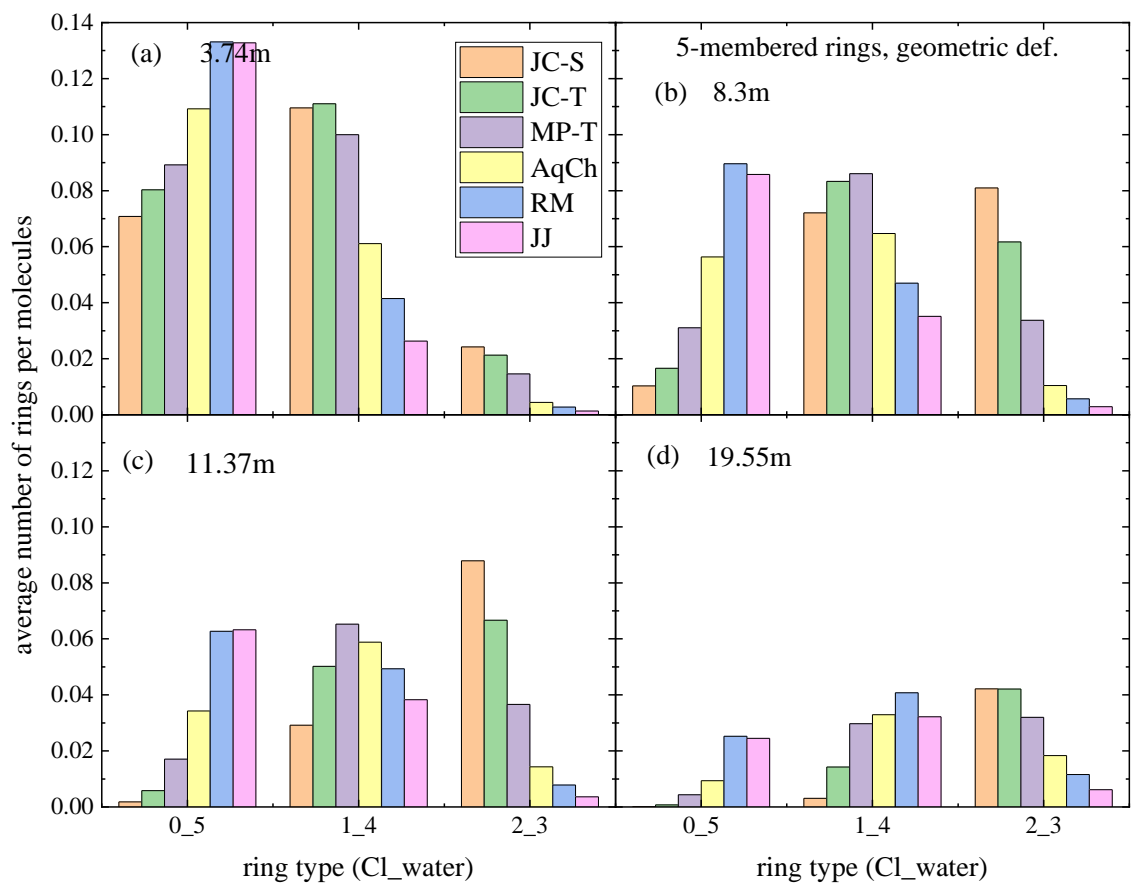

**Figure S15.** Distributions of different types of 5-membered rings (rings contain Cl⁻ ions and water molecules), normalized by the number of molecules (water + Cl⁻ ions), at different concentrations ( (a) 3.74 mol/kg, (b) 8.3 mol/kg, (c) 11.37 mol/kg, (d) 19.55 mol/kg), obtained from different MD models, using the geometric definition of H-bond.

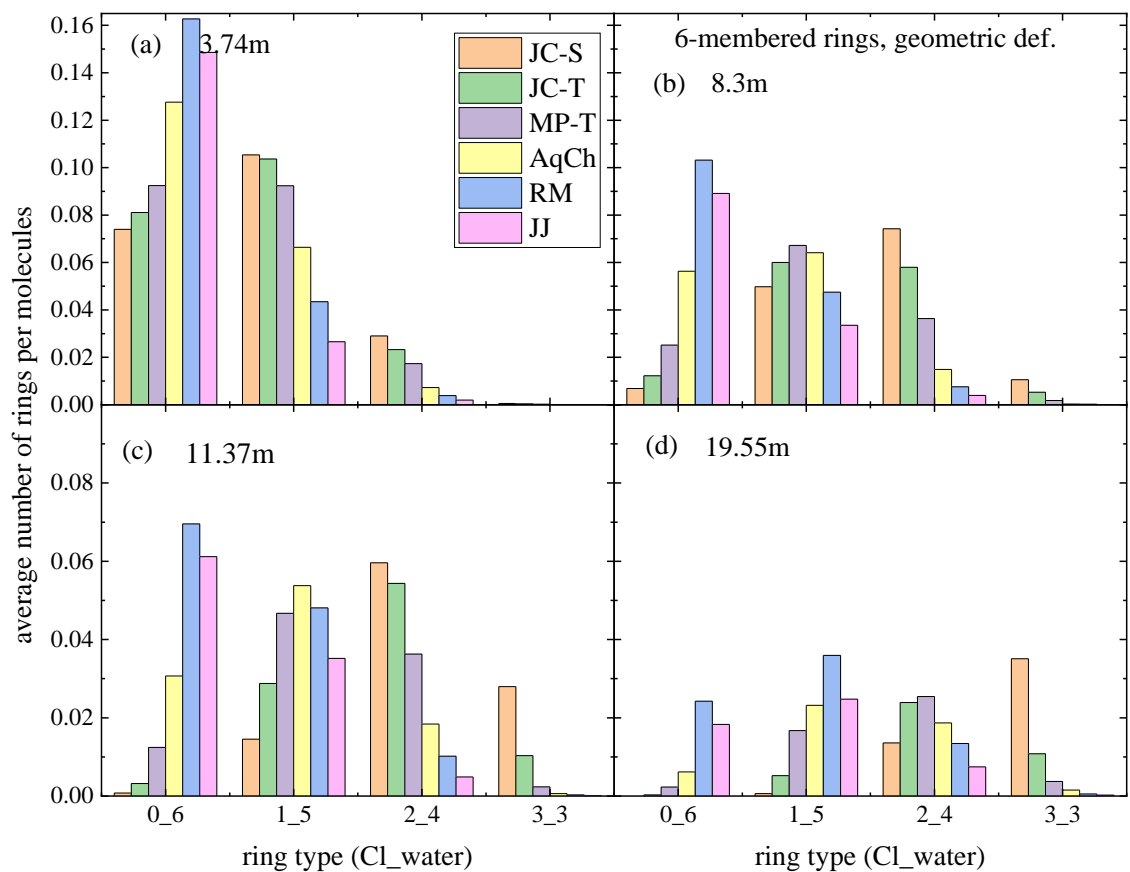

**Figure S16.** Distributions of different types of 6-membered rings (rings contain Cl⁻ ions and water molecules), normalized by the number of molecules (water + Cl⁻ ions), at different concentrations ( (a) 3.74 mol/kg, (b) 8.3 mol/kg, (c) 11.37 mol/kg, (d) 19.55 mol/kg), obtained from different MD models, using the geometric definition of H-bond.

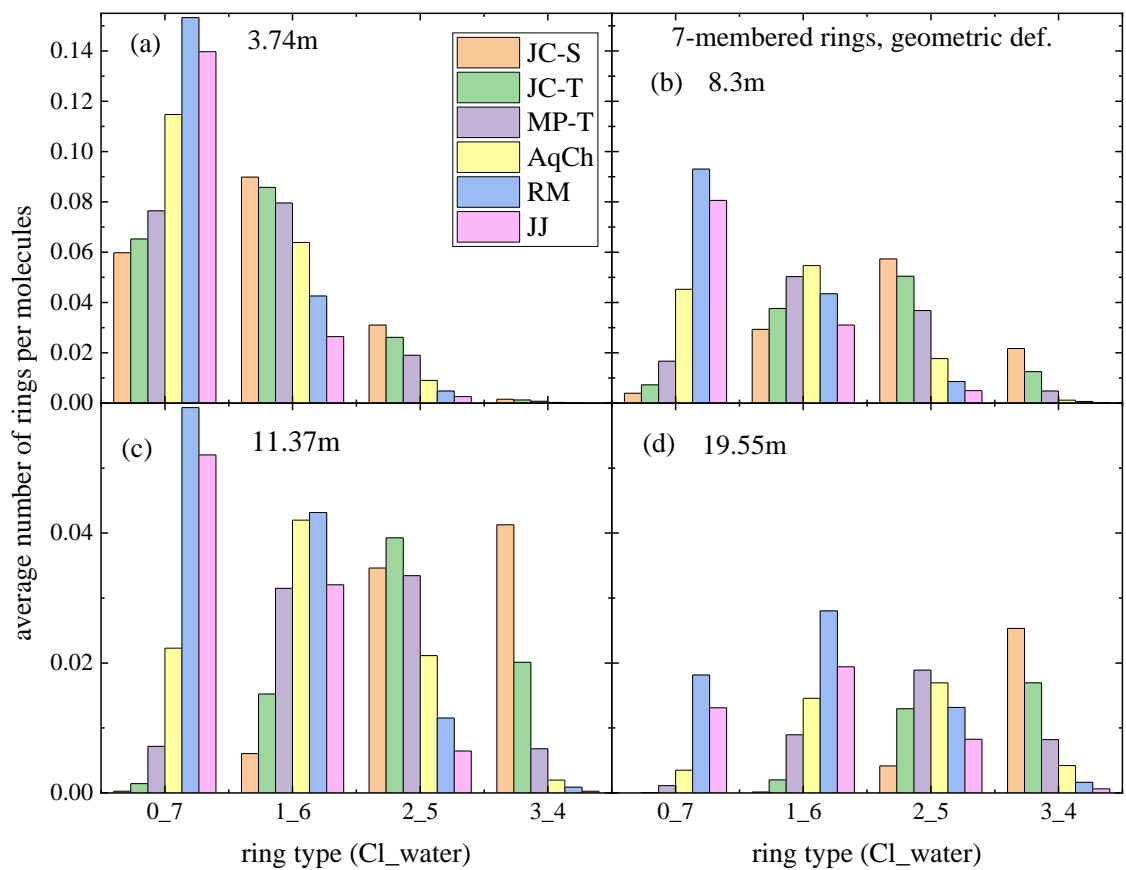

**Figure S17**. Distributions of different type 7-membered rings (rings contain Cl⁻ ions and water molecules), normalized by the number of molecules (water + Cl⁻ ions), at different concentrations ( (a) 3.74 mol/kg, (b) 8.3 mol/kg, (c) 11.37 mol/kg, (d) 19.55 mol/kg), obtained from various MD models, using the geometric definition of H-bond.

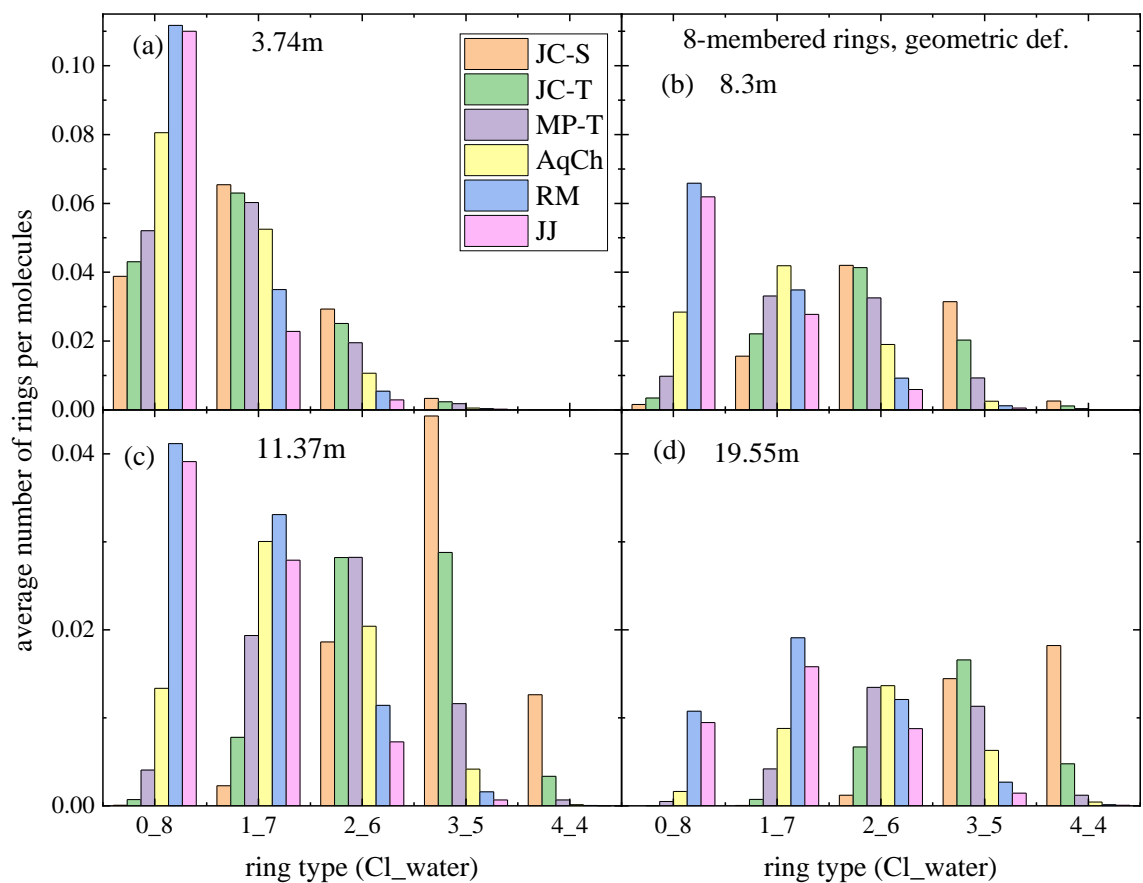

**Figure S18.** Distributions of different types of 8-membered rings (rings contain Cl[-] ions and water molecules), normalized by the number of molecules (water + Cl[-] ions), at different concentrations ( (a) 3.74 mol/kg, (b) 8.3 mol/kg, (c) 11.37 mol/kg, (d) 19.55 mol/kg), obtained from various MD models, using the geometric definition of H-bond.